# Global Smart Spaces

# D11: SECOND SET OF SPACES

EVANGELOS ZIRINTSIS, GRAHAM KIRBY, ALAN DEARLE, BEN ALLEN, ROB MACINNIS, ANDREW MCCARTHY, RON MORRISON, PADDY NIXON, ANDREW JAMIESON, CHRIS NICHOLSON, STEVEN HARRIS

| | | | |
|---|---|---|---|
| **IST Project Number** | IST-2000-26070 | **Acronym** | GLOSS |
| **Full title** | Global Smart Spaces | | |
| **EU Project officer** | Thomas Skordas | | |

| | | | | | | |
|---|---|---|---|---|---|---|
| **Deliverable** | **Number** | D11 | **Name** | | | |
| **Task** | **Number** | T | **Name** | | | |
| **Work Package** | **Number** | WP | **Name** | | | |
| **Date of delivery** | **Contractual** | | | **Actual** | | |
| **Code name** | | | | **Version** 1.0 draft ☐ final ☑ | | |
| **Nature** | Prototype ☐ Report ☑ Specification ☐ Tool ☐ Other: | | | | | |
| **Distribution Type** | Public ☑ Restricted ☐ to: <partners> | | | | | |
| **Authors (Partner)** | University of St Andrews and University of Strathclyde | | | | | |
| **Contact Person** | Graham Kirby | | | | | |
| | **Email** | graham@dcs.st-and.ac.uk | **Phone** | | **Fax** | |
| **Abstract (for dissemination)** | This document describes the Gloss infrastructure supporting implementation of location-aware services. The document is in two part. The first part describes software architecture for the smart space. As described in D8, a **local architecture** provides a framework for constructing Gloss applications, termed **assemblies**, that run on individual physical nodes, whereas a **global architecture** defines an overlay network for linking individual assemblies. The second part outlines the hardware installation for local sensing. This describes the first phase of the installation in Strathclyde University. A installation is planned for Trinity College Dublin – however this is delayed due to manufacturer delays in supplying hardware elements. The Construction guidelines for this hardware are detailed in D12. | | | | | |
| **Keywords** | | | | | | |

# 1 INTRODUCTION

This document describes the Gloss infrastructure supporting implementation of location-aware services. The document is in two part. The first part describes software architecture for the smart space. As described in D8, a **local architecture** provides a framework for constructing Gloss applications, termed **assemblies**, that run on individual physical nodes, whereas a **global architecture** defines an overlay network for linking individual assemblies. The second part outlines the hardware installation for local sensing. This describes the first phase of the installation in Strathclyde University. A installation is planned for Trinity College Dublin – however this is delayed due to manufacturer delays in supplying hardware elements. The Construction guidelines for this hardware are detailed in D12.

# 2 PART I – SOFTWARE INFRASTRUCTURE

The structure of the software is outlined in Figure 1. At the top level, among other packages, the *infrastructure* package contains support for both local and global architectures; *model* contains the Gloss ontology. The local infrastructure package includes: the *assembly* package containing various assemblies; the *component* package which includes implementations of Gloss components; the *factory* package which allows users to create components using factories; the *pipeline* package which provides the interfaces that the Gloss components implement; the *ui* package which provides the User Interface. The *model* package contains implementations of different aspects of the Gloss ontology, such as context, space, time etc.

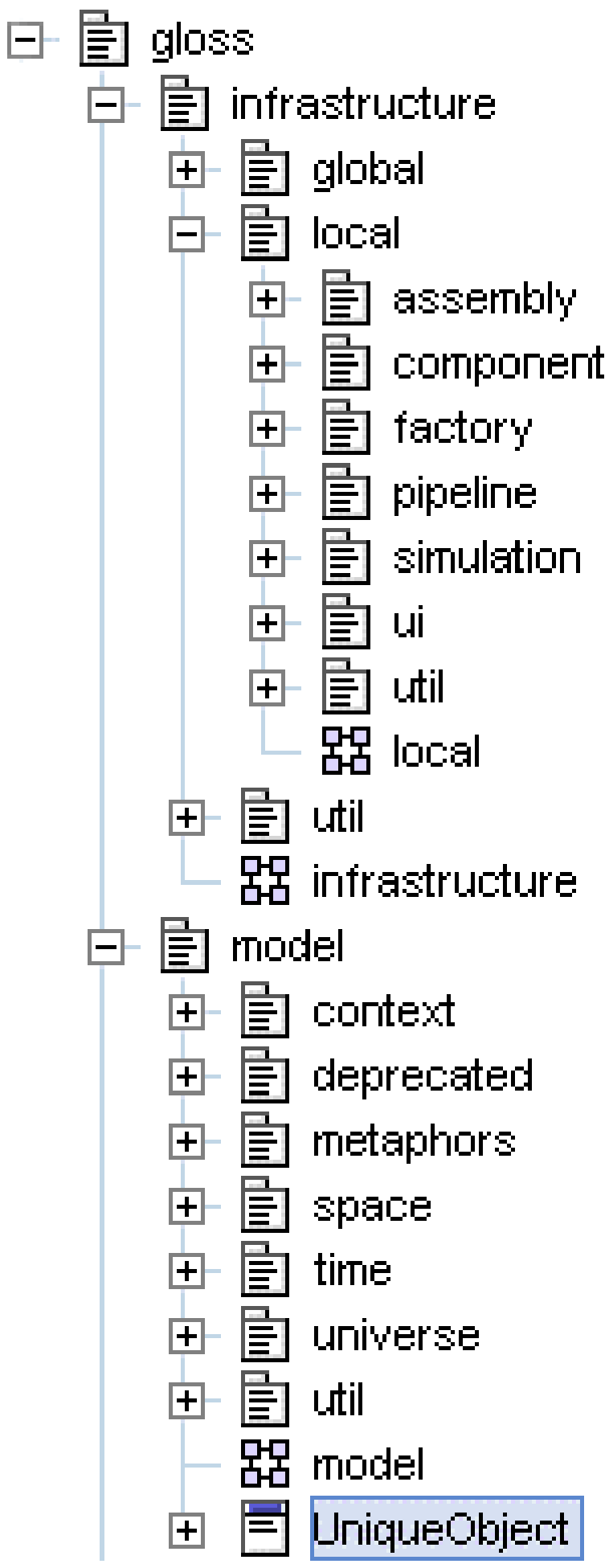

**Figure 1:** Gloss Software Outline

Particular location-aware services, such as *Radar*, *Hearsay* or *Trails*, are implemented as distributed pipelines of communicating assemblies. Each assembly is a pipeline of modular components. Events flow between components as strings, XML fragments or structured objects, as appropriate.

Using the pipeline architectural pattern has a number of advantages:

- **Modularity:** each component is independent, which eases development and maintenance.
- **Flexibility:** the pipeline architecture allows applications to adapt to changing requirements in a flexible manner.
- **Software Reuse:** new applications can easily be constructed using existing software. Each component is independent and can be easily reused in multiple assemblies.
- **Extensibility:** new components may be developed and added without impacting an existing system.

# 3 LOCATION-AWARE SERVICES PIPELINE

The pipeline makes use of both hardware and software components. The hardware components used are shown in Figure 2. A PocketPC PDA connects via Bluetooth to a GPS device and a mobile phone. The mobile phone communicates with the Gloss infrastructure via TCP/IP over GPRS.

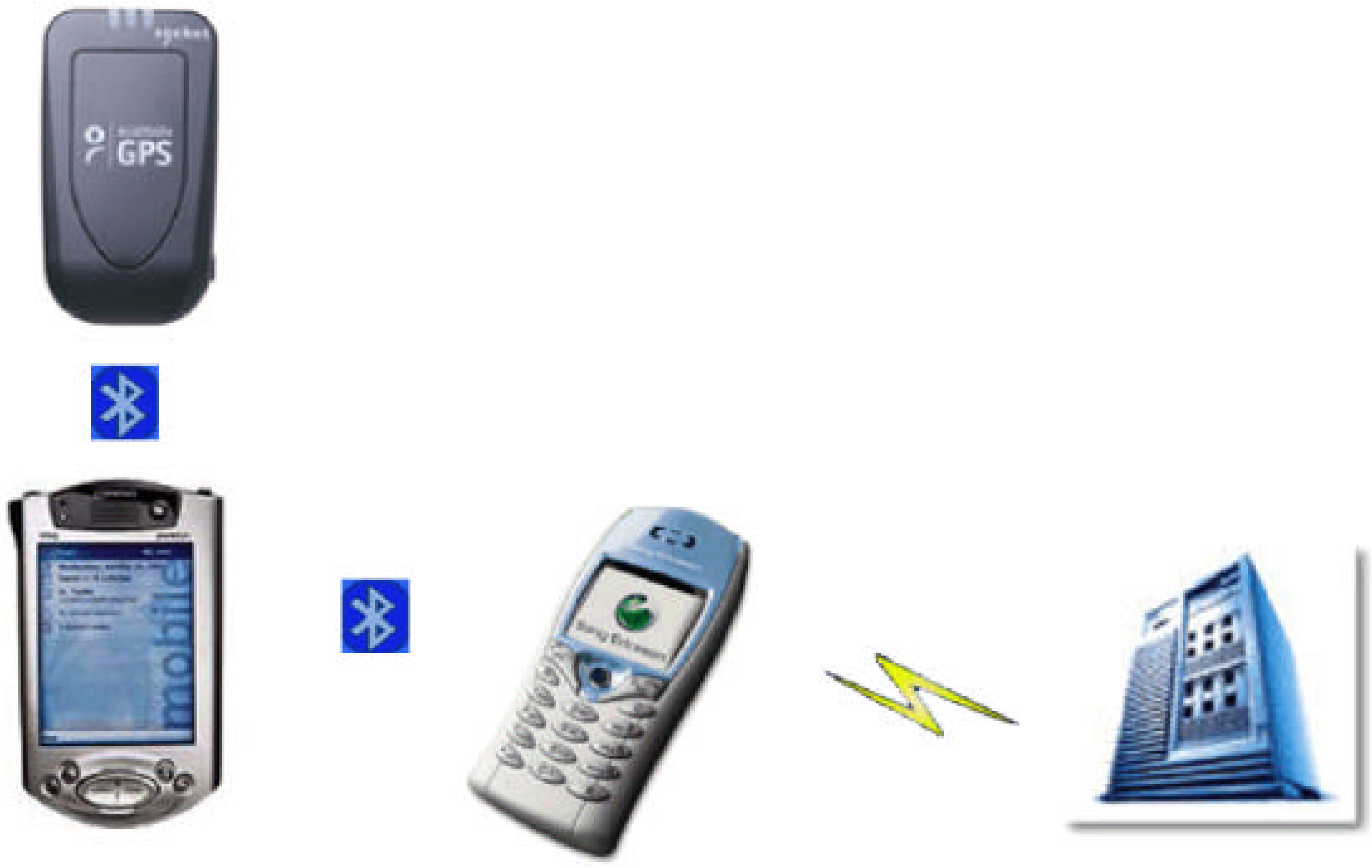

**Figure 2:** Hardware used in the pipeline

In the rest of this section we describe a particular pipeline of software components which are deployed within the hardware infrastructure shown above. Via this pipeline, the user is provided with location-aware services such as Hearsay, Radar and Trails. It consists of two communicating assemblies: the Mobile Application Assembly (MAA) and the Generic Server Assembly (GSA).

## 3.1 MOBILE APPLICATION

Figure 3 shows the MAA in use.

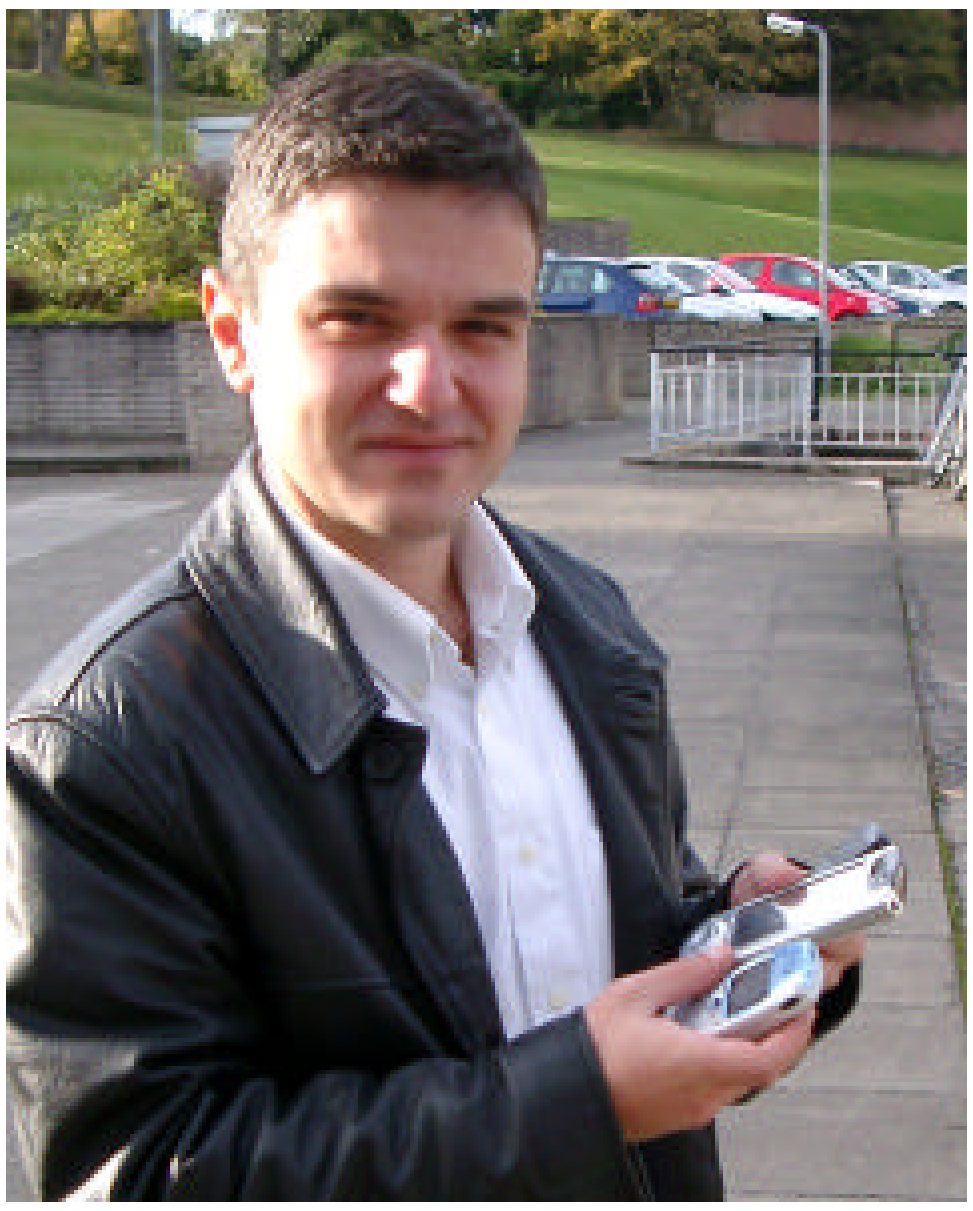

**Figure 3**: Mobile assembly in use

Figure 4 shows the MAA structure; it consists of two assemblies, the User Interface Assembly (UIA) and the GPS Assembly (GPSA).

The UIA receives events from a server and displays the corresponding graphical components within a *Display* area. The *Receiver* processes any remote events and distributes them to the appropriate interface layer. The current implementation provides four different layers: *Trails*, *Radar*, *Hearsay* and *Map*. These receive location data, hearsay events and map data respectively. Each of the layers is displayed within an area specified by component *GUI*. The *Controller* component receives user commands from the *GUI* and performs the appropriate task. The assembly is structured such that it would be straightforward to add or modify the interface layers.

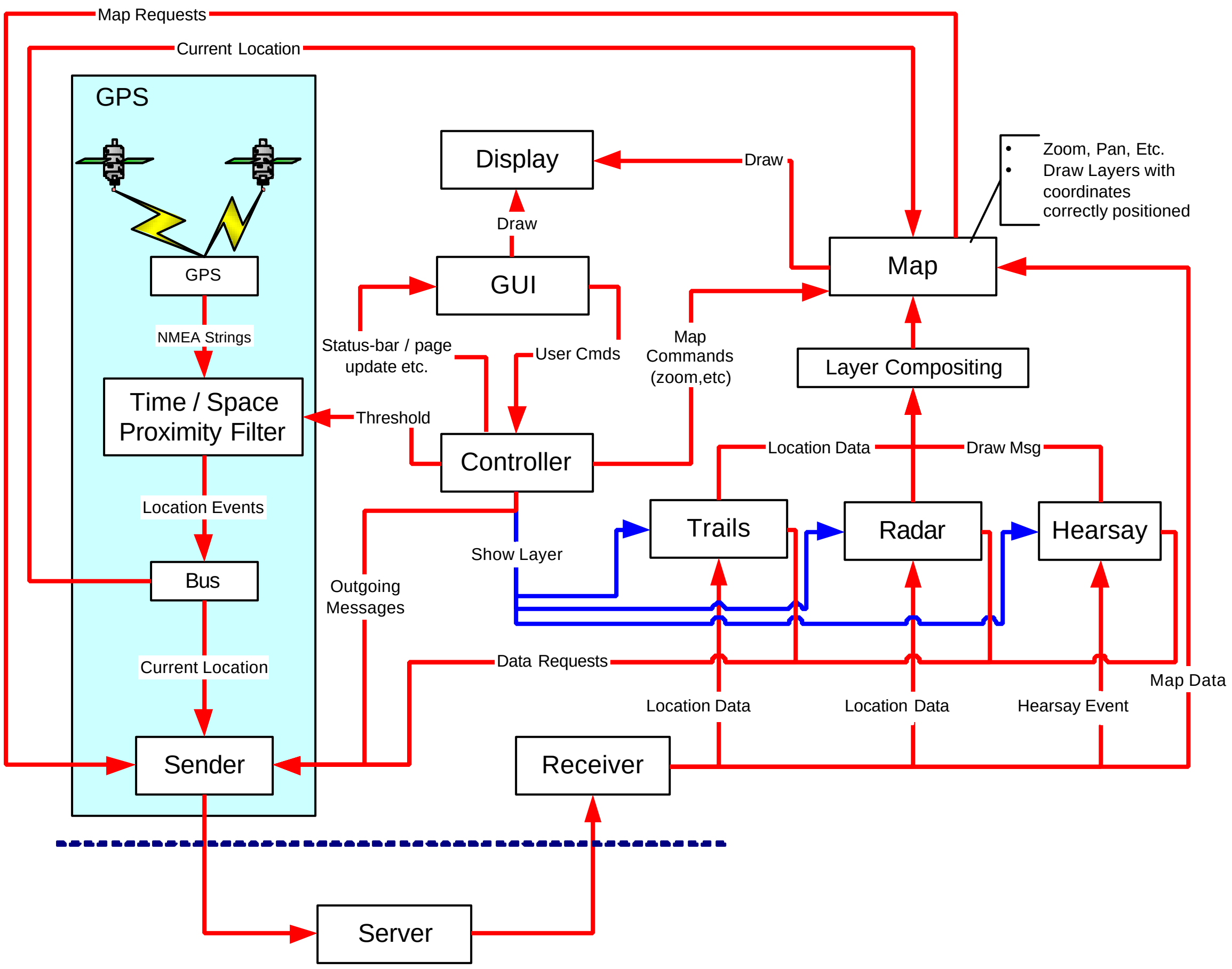

**Figure 4:** The Mobile Application Assembly

The GPS Assembly, within the shaded rectangle in the previous diagram, is shown in more detail in **Figure 5**. The GPS device is a component that establishes a connection with the GPS reader. The readings from the GPS device are in the form of NMEA strings. The *NMEALocationFilter* passes through only those strings containing location information. The *EventBus* forwards these events to multiple components, which convert various device-dependent NMEA strings into *Observation* objects. These are then forwarded to the *ObservationBuffer* component which combines location information with time of observation and GPS meta-data such as the number of satellites visible etc. The *TimeSpaceProximityFilter* filters these objects; only those which differ in time or space from the previous reading by more than appropriate threshold values are injected into the pipeline. These are passed to an *ObjectToXML* adapter for serialisation to XML representation. The final XML message is passed to the *Sender*, which sits on top of various 3rd party components that allow communication with the infrastructure over TCP/IP.

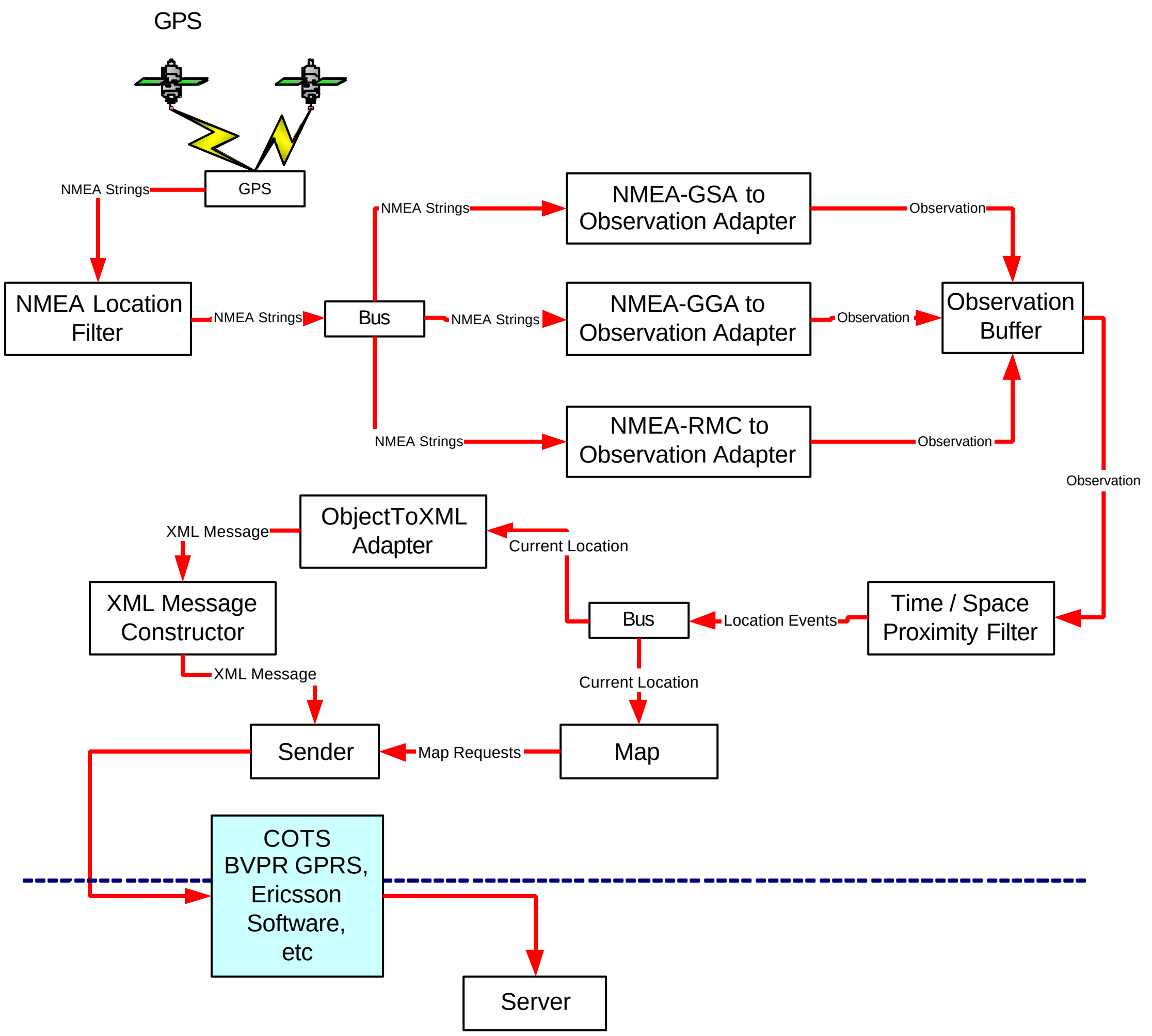

**Figure 5:** The GPS Assembly software structure

## 3.2 GENERIC SERVER

Figure 6 shows the components of the Generic Server Assembly. The *IPSocketAdapter* component receives string events. These are filtered by the *XMLFilter*, which passes through only well-formed XML messages, and a security checker that authenticates the origins of the message. Messages are then forwarded to an *EventBus*, which sends it to five components: another *EventBus* component and four modules which process map, hearsay, radar and trails events (*MapModule*, *HearsayModule*, *RadarModule*, *TrailsModule* respectively). All of these components generate responses in XML format which are forwarded to the *EventServer*, to be sent over the network to the appropriate clients.

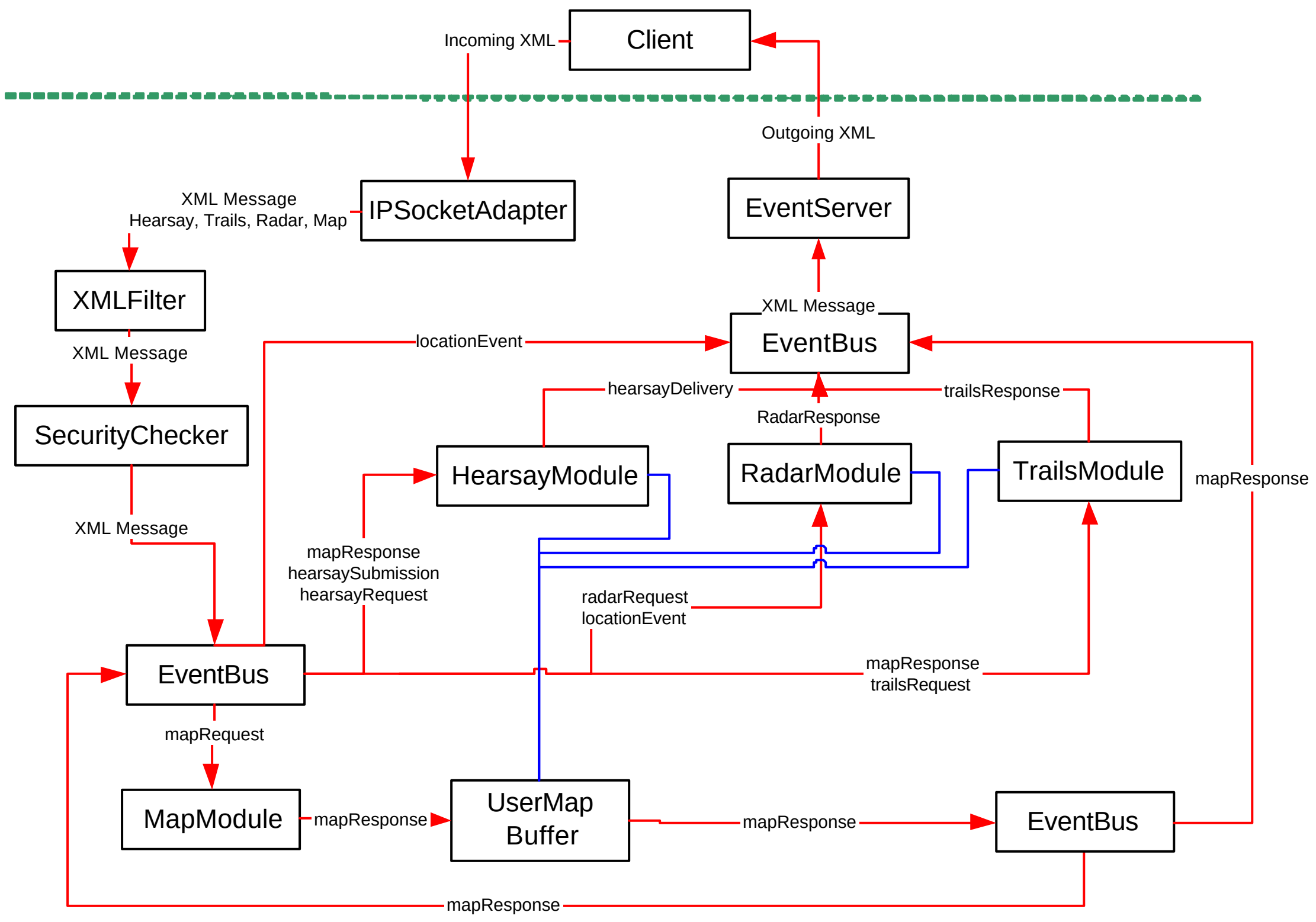

**Figure 6:** The Generic Server Assembly Software structure

The *MapModule* accepts a *mapRequest* event. If there is an appropriate map, the module generates a *mapResponse* which is forwarded to the *UserMapBuffer* component. This caches the current map view of the user specified in the event. The *mapResponse* event is also forwarded to the *HearsayModule*, *RadarModule* and *TrailsModule*, which generate individual appropriate responses.

The *HearsayModule* accepts three kinds of events: *hearsayRequest*, *hearsaySubmission* and *mapResponse*. On a *hearsayRequest* event, the module records whether the user specified by the ID tag in the message wishes to receive hearsay messages. On a *hearsaySubmission* event, the module matches the specified hearsay delivery context against the context of all current users, and generates a *hearsayDelivery* event for each match. On a *mapResponse* event, the module is notified that a user's view has changed, in which case it sends to the user, via *EventServer*, all the previously recorded *hearsayDelivery* messages that are within the new view. The information about the user's view is retrieved from the *UserMapBuffer* component.

The *RadarModule* accepts two kinds of events: *radarRequest* and *locationEvent*. On a *radarRequest* event, the module records whether the user specified by the ID tag in the message wishes to receive radar messages. On a *locationEvent* event, the module finds the users whose current view contains the coordinates specified by the *locationEvent*. It then generates *radarResponse* messages for these users and forwards them to the *EventServer*.

The *TrailsModule* accepts three kinds of events: *trailsRequest*, *trailSubmission* and *mapResponse*. On a *trailsRequest* event, the module records whether the user specified by the ID tag in the message wishes to receive trails messages. The module also records the IDs of the users that the particular client might be interested to receive trails from. If there are no users then this means that the client is interested in all trails. On a *trailsSubmission* event, the module records the locations of trails included. On a *mapResponse* event, the module is notified that a user's view has changed, in which case it sends to the user, via *EventServer*, all the previously recorded trails that are within the new view.

All the response messages generated by the different modules contain an ID tag that specifies the user. The user information is used by the *EventServer* which filters the messages and sends them to the appropriate users.

### 3.2.1 EVENT TYPES

Several kinds of events have been mentioned in the previous section. In this section, we give examples for each of those events.

#### 3.2.1.1 LOCATIONEVENT

Location event messages are defined by the schema located at:

http://www-systems.dcs.st-and.ac.uk/gloss/xml/2003-07/locationEvent.xsd

An example location event is shown below:

```
<locationEvent>
    <ID>
            <email>vangelis@dcs.st-and.ac.uk</email>
    </ID>
    <processingSequence />
    <observation>
            <timeOfObservation>2003-8-17T18:31:59:516</timeOfObservation>
            <where>
                    <physicalLocation>
                            <coordinate>
                                    <latLongCoordinate>
                                            <latitude>56.340232849121094</latitude>
                                            <longitude>-2.808</longitude>
                                    </latLongCoordinate>
                            </coordinate>
                    </physicalLocation>
            </where>
    </observation>
</locationEvent>
```

#### 3.2.1.2 HEARSAYREQUEST

Hearsay request contains the ID of the user and whether the service should be activated or not. An example is shown below:

```
<hearsayRequest>
    <ID>
            <email>graham@dcs.st-and.ac.uk</email>
    </ID>
    <activate>true</ activate >
</ hearsayRequest >
```

#### 3.2.1.3 HEARSAYSUBMISSION

Hearsay submission contains location information about the sender and the receiver as well as the hearsay message itself. An example is shown below:

```
<hearsaySubmission>
    <sender>
```

```
            <locationEvent>
                <ID>
                    <email>al@dcs.st-and.ac.uk</email>
                </ID>
                <processingSequence />
                <observation>
                    <timeOfObservation>2003-05-16T18:31:59:516</timeOfObservation>
                    <where>
                        <physicalLocation>
                            <coordinate>
                                <latLongCoordinate>
                                    <latitude>56.360232849121094</latitude>
                                    <longitude>-2.80704378657099</longitude>
                                </latLongCoordinate>
                            </coordinate>
                        </physicalLocation>
                    </where>
                </observation>
            </locationEvent>
        </sender>
        <receiver>
            <locationEvent>
                <ID>
                    <email>ron@dcs.st-and.ac.uk</email>
                </ID>
                <processingSequence />
                <observation>
                    <timeOfObservation>2003-8-17T18:31:59:516</timeOfObservation>
                    <where>
                    <physicalLocation>
                        <coordinate>
                            <latLongCoordinate>
                                <latitude>56.340232849121094</latitude>
                                <longitude>-2.808</longitude>
                            </latLongCoordinate>
                        </coordinate>
                    </physicalLocation>
                    </where>
                </observation>
            </locationEvent>
        </receiver>
        <hearsayMessage>Hello Vangelis</hearsayMessage>
    </hearsaySubmission>
```

#### *3.2.1.4 HEARSAYDELIVERY*

Hearsay delivery contains location information about the sender and the receiver as well as the hearsay message itself. An example is shown below:

```
    <hearsayDelivery>
        <ID>
            <email>rob@dcs.st-and.ac.uk</email>
        </ID>
        <sender>
            <locationEvent>
                <ID>
                    <email>al@dcs.st-and.ac.uk</email>
                </ID>
                <processingSequence />
                <observation>
                    <timeOfObservation>2003-05-16T18:31:59:516</timeOfObservation>
```

```
                        <where>
                            <physicalLocation>
                                <coordinate>
                                    <latLongCoordinate>
                                        <latitude>56.360232849121094</latitude>
                                        <longitude>-2.80704378657099</longitude>
                                    </latLongCoordinate>
                                </coordinate>
                            </physicalLocation>
                        </where>
                    </observation>
                </locationEvent>
            </sender>
            <receiver>
                <locationEvent>
                    <ID>
                        <email>rob@dcs.st-and.ac.uk</email>
                    </ID>
                    <processingSequence />
                    <observation>
                        <timeOfObservation>2003-8-17T18:31:59:516</timeOfObservation>
                        <where>
                            <physicalLocation>
                                <coordinate>
                                    <latLongCoordinate>
                                        <latitude>56.340232849121094</latitude>
                                        <longitude>-2.808</longitude>
                                    </latLongCoordinate>
                                </coordinate>
                            </physicalLocation>
                        </where>
                    </observation>
                </locationEvent>
            </receiver>
            <hearsayMessage>Hello Vangelis</hearsayMessage>
        </hearsayDelivery >
```

### 3.2.1.5 RADARREQUEST

A radar request contains the ID of a user and whether the service should be activated or not. An example is shown below:

```
<radarRequest>
    <ID>
        <email>graham@dcs.st-and.ac.uk</email>
    </ID>
    <activate>false</activate>
</radarRequest>
```

### 3.2.1.6 RADARRESPONSE

A radar response contains location information about a user. An example is shown below:

```
<radarResponse>
    <ID>
        <email>vangelis@dcs.st-and.ac.uk</email>
    </ID>
    <locationEvent>
        <ID>
            <email>al@dcs.st-and.ac.uk</email>
        </ID>
```

```
            <processingSequence />
            <observation>
                  <timeOfObservation>2003-05-16T18:31:59:516</timeOfObservation>
                  <where>
                        <physicalLocation>
                              <coordinate>
                                    <latLongCoordinate>
                                          <latitude>56.360232849121094</latitude>
                                          <longitude>-2.80704378657099878</longitude>
                                    </latLongCoordinate>
                              </coordinate>
                        </physicalLocation>
                  </where>
            </observation>
      </locationEvent>
</radarResponse>
```

#### 3.2.1.7 TRAILREQUEST

A trail request contains the ID of a user and whether the service should be activated or not. Optionally it includes a list of IDs that correspond to the users that it is desired to receive trails from. An example is shown below:

```
<trailRequest>
   <ID>
         <email>al@dcs.st-and.ac.uk</email>
   </ID>
   <activate>true</activate>
   <desiredUsers>
         <ID> <email>vangelis@dcs.st-and.ac.uk</email> </ID>
         <ID> <email>graham@dcs.st-and.ac.uk</email> </ID>
         <ID> <email>ron@dcs.st-and.ac.uk</email> </ID>
         <ID> <email>rob@dcs.st-and.ac.uk</email> </ID>
   </desiredUsers>
</trailRequest>
```

#### 3.2.1.8 TRAILSSUBMISSION

A trail submission contains a sequence of locations. An example is shown below:

```
<trailSubmission>
   <trail>
         <observedTrail>
               <locationEvent>
                  <ID>
                        <email>al@dcs.st-and.ac.uk</email>
                  </ID>
                  <processingSequence />
                  <observation>
                        <timeOfObservation>2003-05-16T18:31:59:516</timeOfObservation>
                        <where>
                              <physicalLocation>
                                    <coordinate>
                                          <latLongCoordinate>
                                                <latitude>56.370232849121094</latitude>
                                                <longitude>-2.80804378657099</longitude>
                                          </latLongCoordinate>
                                    </coordinate>
                              </physicalLocation>
                        </where>
                  </observation>
                  </locationEvent>
```

```
            <locationEvent>
                <ID>
                    <email>al@dcs.st-and.ac.uk</email>
                </ID>
                <processingSequence />
                <observation>
                    <timeOfObservation>2003-05-16T18:32:04:516</timeOfObservation>
                    <where>
                        <physicalLocation>
                            <coordinate>
                                <latLongCoordinate>
                                    <latitude>56.370232849121094</latitude>
                                    <longitude>-2.80804378657099</longitude>
                                </latLongCoordinate>
                            </coordinate>
                        </physicalLocation>
                    </where>
                </observation>
        </locationEvent>
        </observedTrail>
    </trail>
</trailSubmission>
```

### 3.2.1.9 *TRAILSRESPONSE*

A trail response contains trail information about a user. An example of such a response containing observed trails is shown below:

```
<trailsResponse>
    <ID>
        <email>vangelis@dcs.st-and.ac.uk</email>
    </ID>
    <trail>
        <observedTrail>
            <locationEvent>
                <ID>
                    <email>al@dcs.st-and.ac.uk</email>
                </ID>
                <processingSequence />
                <observation>
                    <timeOfObservation>2003-05-16T18:31:59:516</timeOfObservation>
                    <where>
                        <physicalLocation>
                            <coordinate>
                                <latLongCoordinate>
                                    <latitude>56.370232849121094</latitude>
                                    <longitude>-2.80804378657099</longitude>
                                </latLongCoordinate>
                            </coordinate>
                        </physicalLocation>
                    </where>
                </observation>
                </locationEvent>
            <locationEvent>
                <ID>
                    <email>al@dcs.st-and.ac.uk</email>
                </ID>
                <processingSequence />
                <observation>
                    <timeOfObservation>2003-05-16T18:32:04:516</timeOfObservation>
                    <where>
```

```
                              <physicalLocation>
                                   <coordinate>
                                        <latLongCoordinate>
                                             <latitude>56.370232849121094</latitude>
                                             <longitude>-2.80804378657099</longitude>
                                        </latLongCoordinate>
                                   </coordinate>
                              </physicalLocation>
                         </where>
                    </observation>
                    </locationEvent>
               </observedTrail >
          </trail>
</trailsResponse>
```

### *3.2.1.10 MAPREQUEST*

A map request contains the ID of a user, their current location and the desired zoom level. An example is shown below:

```
<mapRequest>
   <ID>
         <email>vangelis@dcs.st-and.ac.uk</email>
   </ID>
   <coordinate>
         <latLongCoordinate>
               <latitude>56.340232849121094</latitude>
               <longitude>-2.808</longitude>
         </latLongCoordinate>
   </coordinate>
   <zoom>5</zoom>
</mapRequest>
```

### *3.2.1.11 MAPRESPONSE*

A map response contains the ID of a user and the information necessary for them to download an image and display it correctly on the screen. The latter information involves: the web location, the width/height of the image, the coordinates of the top left and bottom right corners, the width/height ratio and the zoom level. An example is shown below:

```
<mapResponse>
   <ID>
         <email>vangelis@dcs.st-and.ac.uk</email>
   </ID>
   <image>
         <url>http://www-systems.dcs.st-and.ac.uk:8180/gloss/standrews_city_600600.jpg</url>
         <imageWidth>600</imageWidth>
         <imageHeight>600</imageHeight>
         <corners>
               <topLeft>
                     <latitude>56.370100</latitude>
                     <longitude>-2.842174</longitude>
               </topLeft>
               <bottomRight>
                     <latitude>56.316349</latitude>
                     <longitude>-2.744143</longitude>
               </bottomRight>
         </corners>
         <ratio>
               <widthRatio>1</widthRatio>
```

```
            &lt;heightRatio&gt;1&lt;/heightRatio&gt;
        &lt;/ratio&gt;
        &lt;zoom&gt;5&lt;/zoom&gt;
    &lt;/image&gt;
&lt;/mapResponse&gt;
```

## 3.3 USER INTERFACE

In this section we illustrate the user interface of the running assemblies. Each assembly is an instantiation of the local architecture. Figure 7 shows the result of starting the server assembly.

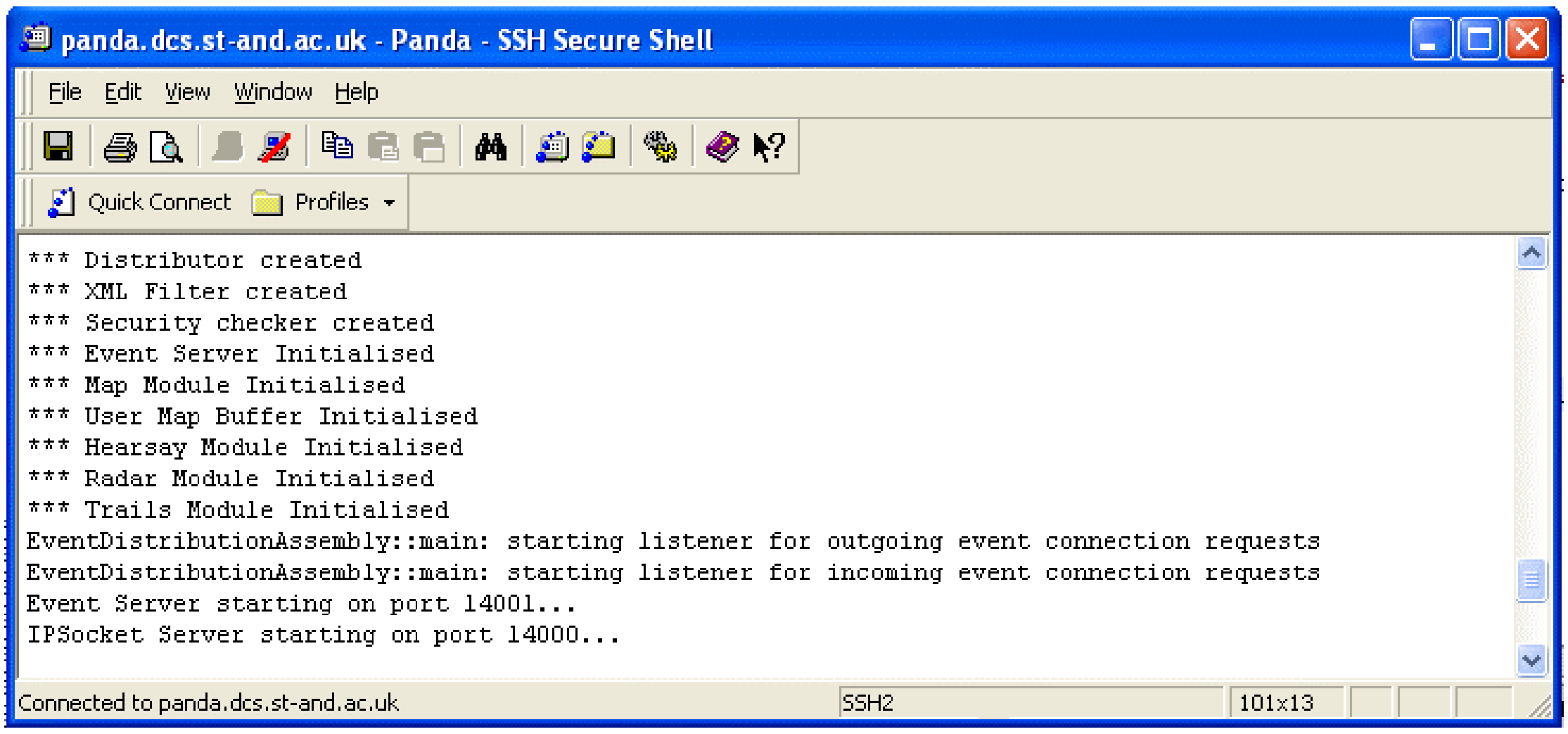

**Figure 7:** Starting the Generic Server Assembly

Figure 9 shows a mobile client in use, while Figure 9 shows a detailed screenshot with the radar service activated. The green circled arrow denotes the current location of the user, and the yellow arrow denotes the location of another user of interest.

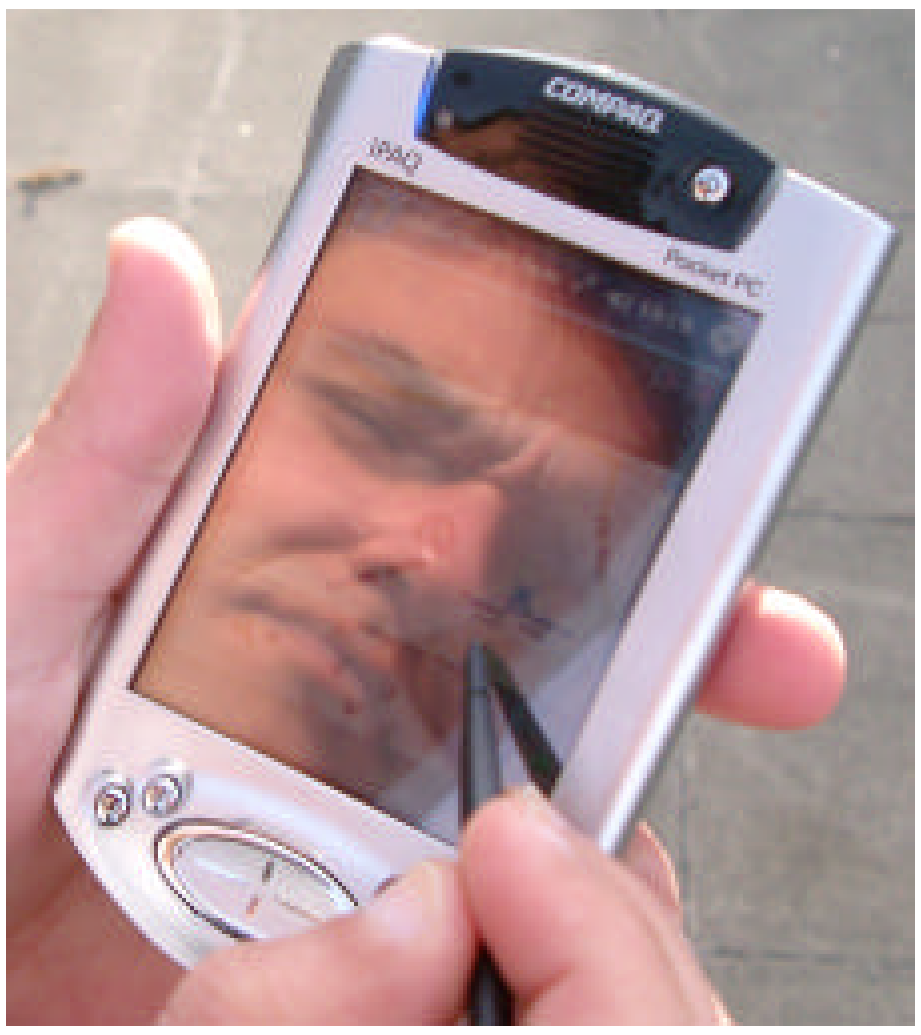

**Figure 8:** Mobile assembly in actual use

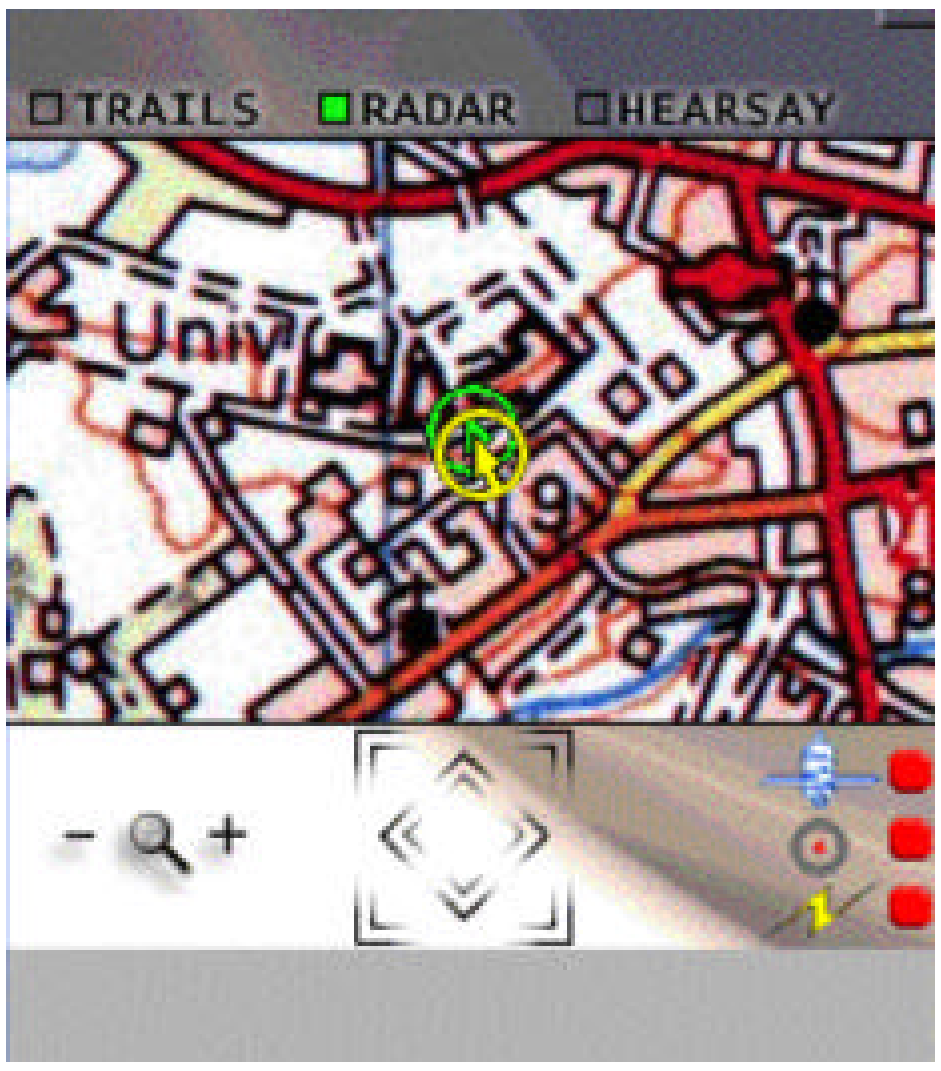

**Figure 9:** A client with radar information

Figure 10 shows the reception of a hearsay message.

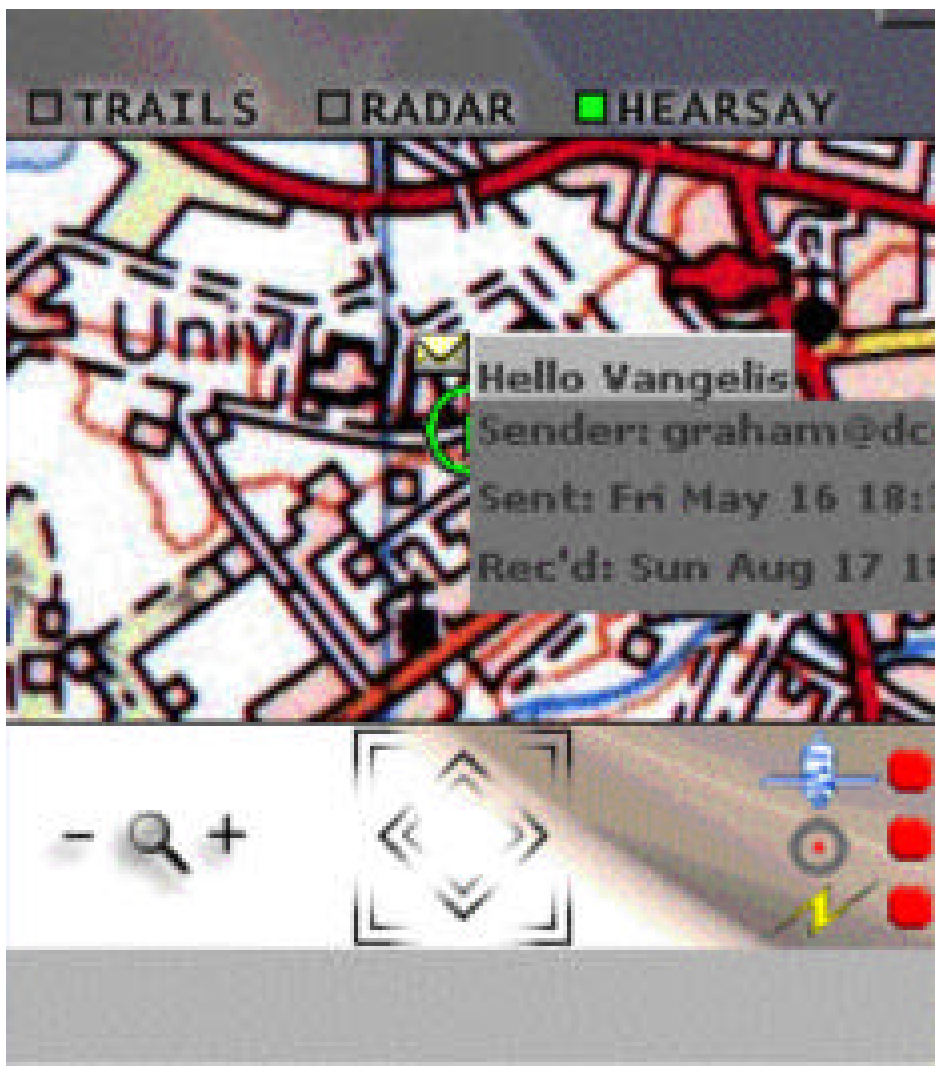

**Figure 10:** A client with hearsay information

**Figure 11** shows the display of both radar and trails information.

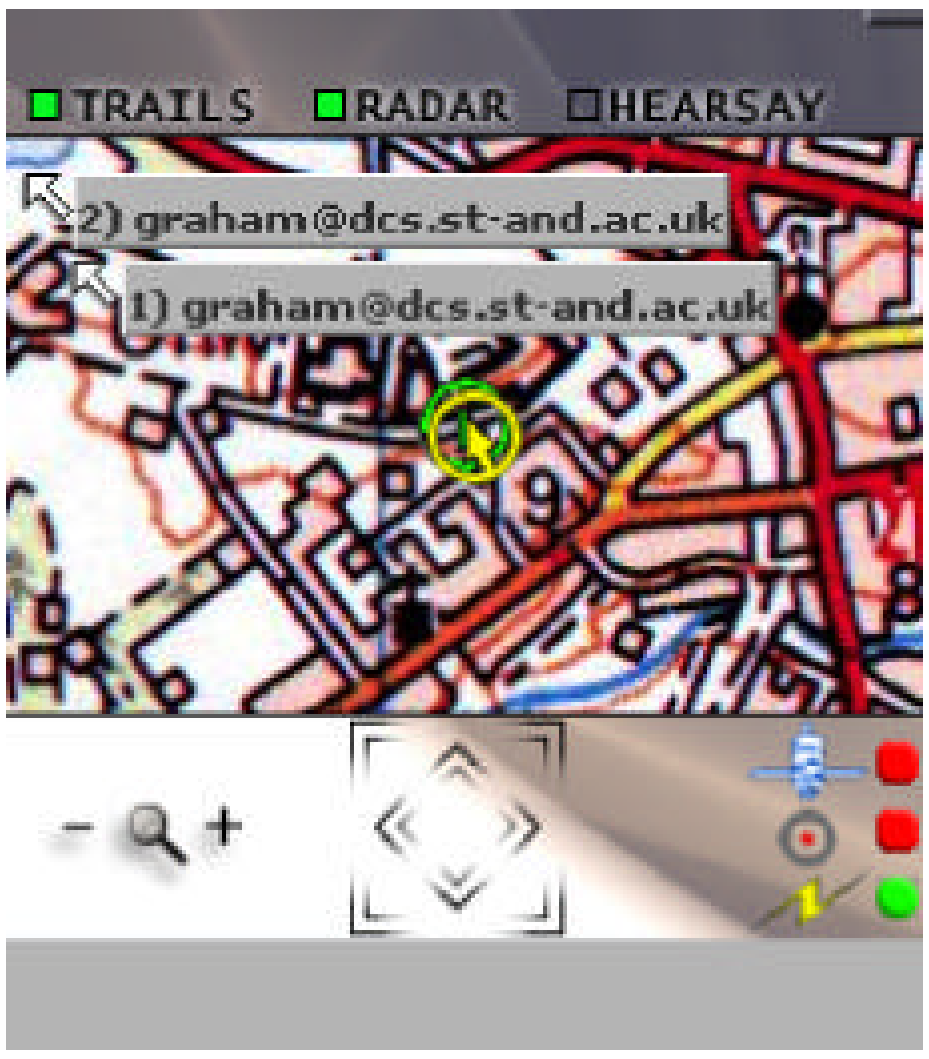

**Figure 11:** A client with radar and trails information

## 3.4 DEPLOYMENT TECHNOLOGY

Global Ubiquitous computing environments require a large and diverse range of inter-communicating services to be deployed at geographically appropriate locations to support their users. Constantly changing requirements and usage patterns necessitate the ability to introduce new components and change – at runtime – the topology and composition of this environment. To start to address these requirements we have developed a deployment engine and a set of associated tools.

### *3.4.1 OVERVIEW*

We introduce several concepts which form the core of the Deployment Engine: Mobile Code Tools and XML Control Documents.

Core components of the deployment process are mobile code tools and XML Control Documents – “To Do Lists” and “Task Reports”. To Do Lists are composed of a set of Tasks which detail actions a tool must attempt to perform upon arrival at a Thin Server. Consequent Task Report documents list the outcomes of each Task and any other associated information. When the tool completes its assigned Tasks a Task Report is sent back to the Deployment Engine. An example To Do list and Task Report are shown below:

```
<ToDoList>
   <Task guid="urn:gloss:aEcncdeEe" type="INSTALL">
        <datum id="PayloadRef">urn:gloss:a222jdjd2s</datum>
   </Task>
   <Task guid="urn:gloss:aBcbcdebe" type="INSTALL">
        <datum id="PayloadRef">urn:gloss:b333jdjd2s</datum>
   </Task>
</ToDoList>

<TaskReport>
   <TaskOutcome guid="urn:gloss:aEcncdeEe" success="TRUE">
        <!-- TaskOutcomes can have zero, one or many datum elements
        which are bindings and data this permits any application
        specific information to be sent back to the Deployment
        Engine -->
        <datum id="StoreGuid">AECJCJDKSKDLDJSUVDJD</datum>
   </TaskOutcome>
   <TaskOutcome guid="urn:gloss:aBcbcdebe" success="FALSE">
        <datum id="Error">403</datum>
   </TaskOutcome>
</TaskReport>
```

Mobile Code Tools are CINGAL bundles which are configurable by attaching an appropriate To Do List to the bundle which encloses the Tool. The deployment engine utilises three primary tools: *Installers*, *Runners* and *Wirers*. Installer Tools install an arbitrary number of bundles into the store of the Thin Server they are sent to, Runner Tools start the execution of a bundle which is already in the store of a Thin Server and Wirer Tools are responsible for making concrete connections between pairs of Abstract Channels.

### *3.4.2 DEPLOYMENT METHOD*

The Deployment Engine distributes autonomous components which perform a specific computation/function (service). The components have no knowledge of the topology of the network of which they are a part. The system takes a deployment specification and from this, deploys the components, starts them running and finally connects them into the specified topology.

Deployment States:

- **Deployed** – corresponds to the state when all bundles have been installed into the TSStore of their respective nodes.
- **Running** – corresponds to the state when all bundles have started computation. Any read/write on abstract channels will block as they have not been connected to a transport mechanism at this state.
- **Wired** – corresponds to the state when all bundles have started computation and all abstract channels have been connected to a transport mechanism which will carry data to their appropriate destination.

A Deployment Descriptor Document (DDD) is a static description of a distributed graph of components. The DDD specifies where to retrieve components (Bundles), machines available, mapping of components to machines (a deployment) and the connections between abstract channel pairs.

Deployment Descriptor Document Multiplicities:

- **Node** – a physical machine is a node.
- **Machine** – a virtual machine running a bundle, many machines run on a node. One bundle is running in each machine.

Figure 12 shows an example Deployment Description describing a GLOSS infrastructure.

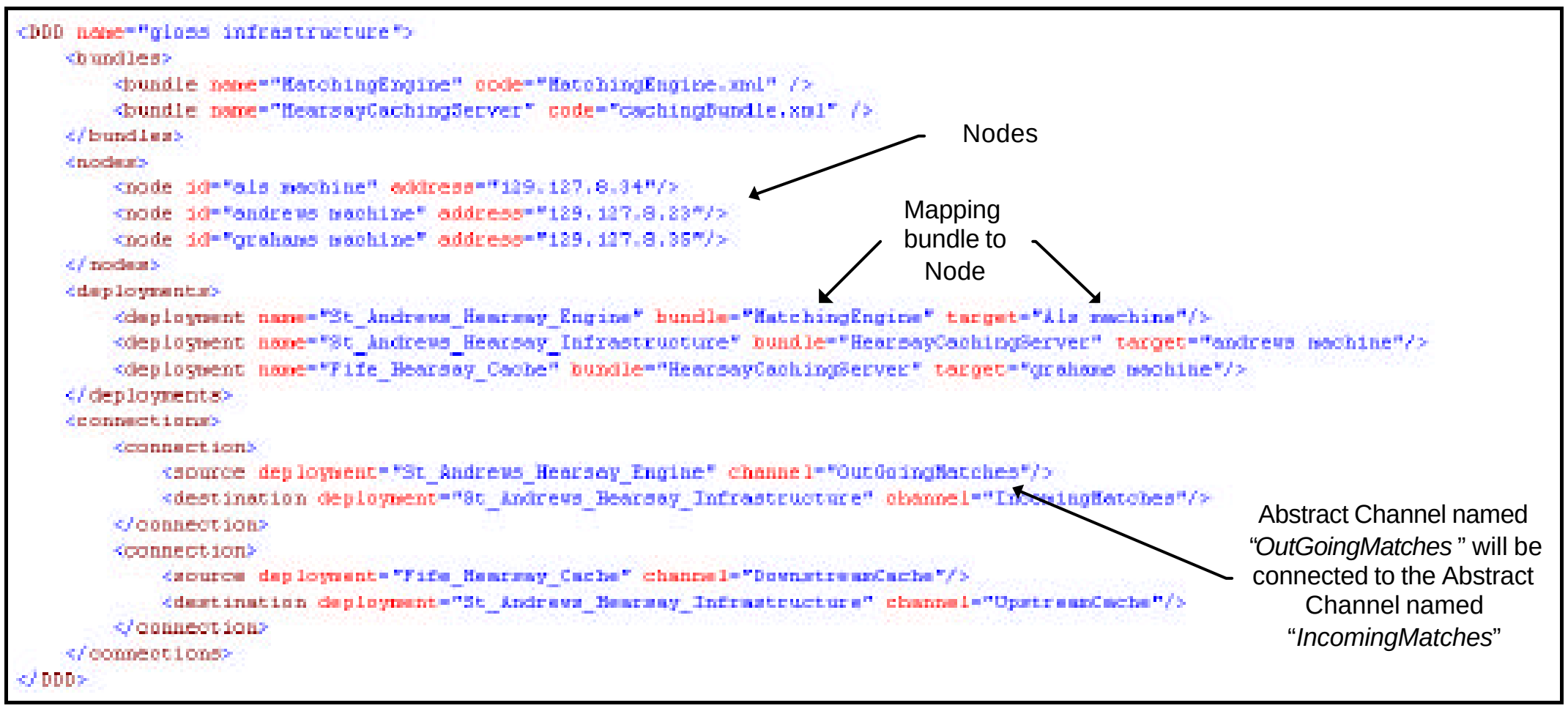

Figure 12: Example DDD

The DDD is input to the Deployment Engine (this process is known as compilation of the DDD), which retrieves bundles from a component catalogue and performs appropriate configuration and firing of a set of Installers, Runners and Wirers to construct and activate (run the connected graph of bundles) the graph described in the DDD.

### *3.4.3 COMPILATION OF DDD*

Installers are configured (by creating an appropriate To Do List) and fired (sent to appropriate nodes and executed) to install required components onto Thin Servers throughout the network. One installer is fired per Thin Server. Each installer sends back a report to the deployment engine listing the TSGUID each bundle has been installed as. Figure 13 shows an example of installation.

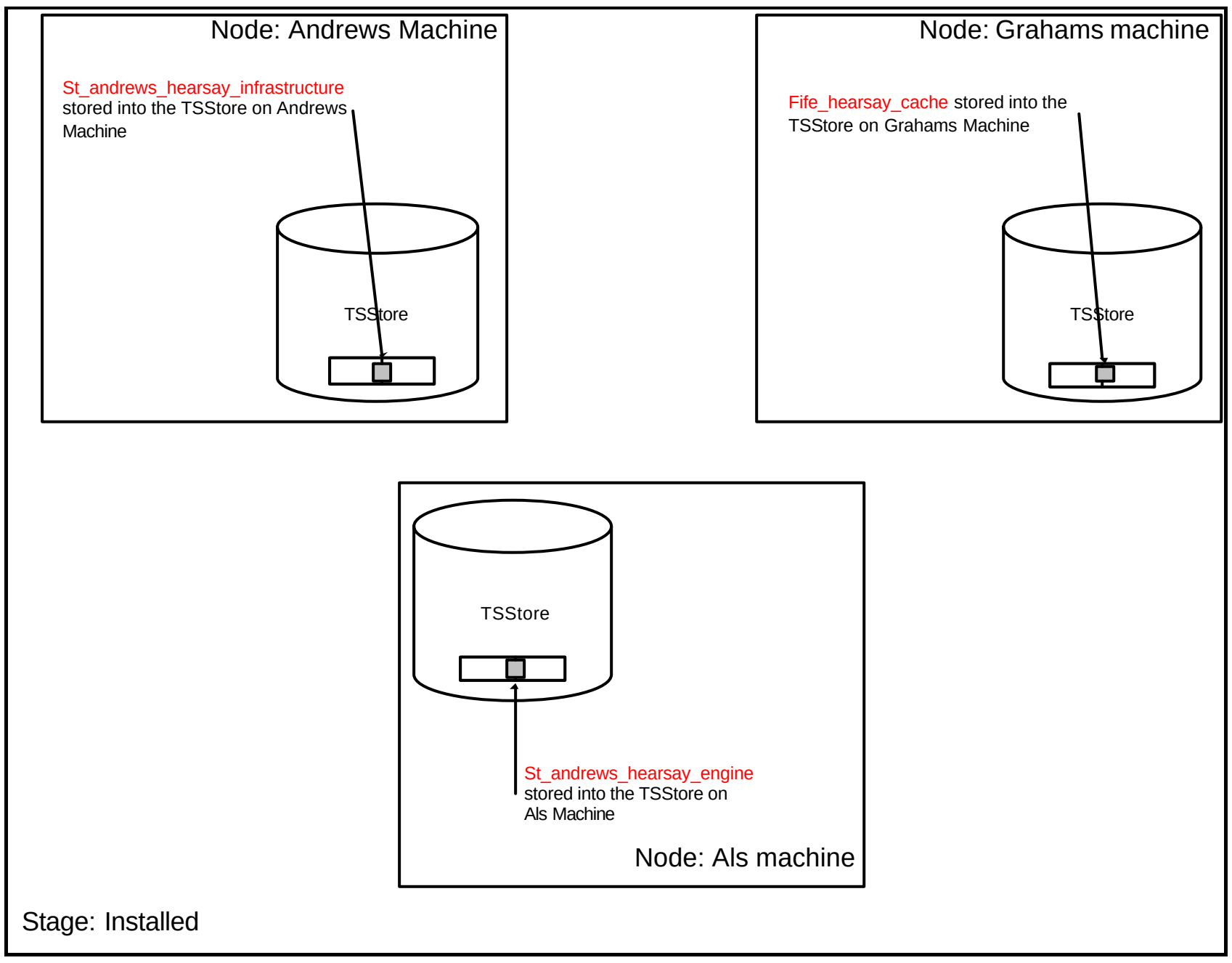

Figure 13: Example of installation

Runners are configured (by creating an appropriate To Do List) and fired to start execution of all 'dormant' installed bundles for this deployment. One runner is fired per Thin Server. Each runner sends back a report to the deployment engine with a serialized TSConnector for each bundle fired. Figure 14 shows an example of running.

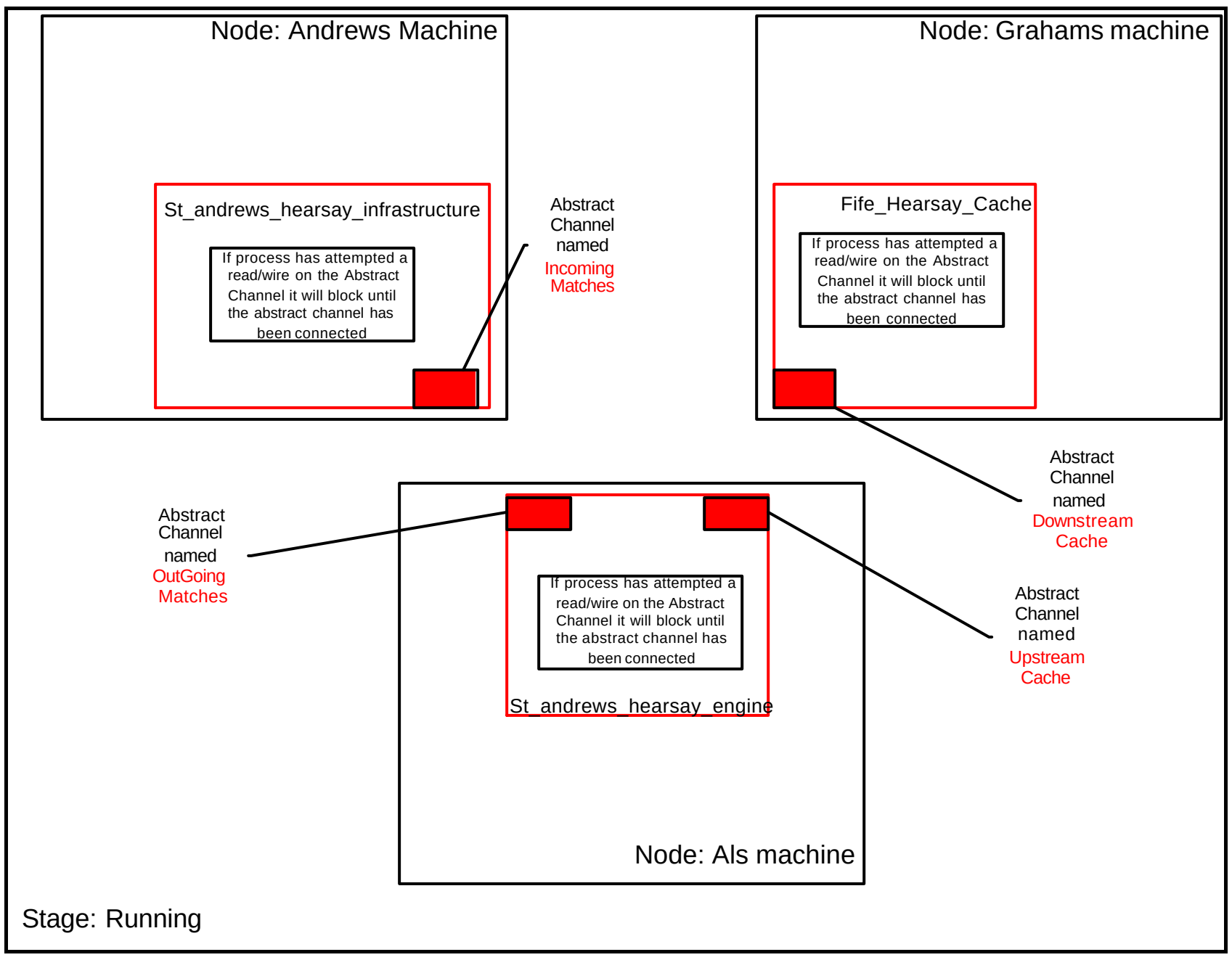

Figure 14: Example of running

Wirers are configured (by creating an appropriate ToDoList) and fired to connect abstract channels in each machine. One wirer is used per connection. The wirer is sent to one Thin Server which requires a connection or maintenance and produces

'offspring' (other wirer bundles) which are sent to other nodes to complete the operation of connecting Abstract Channels as required. Figure 15 shows an example of wiring.

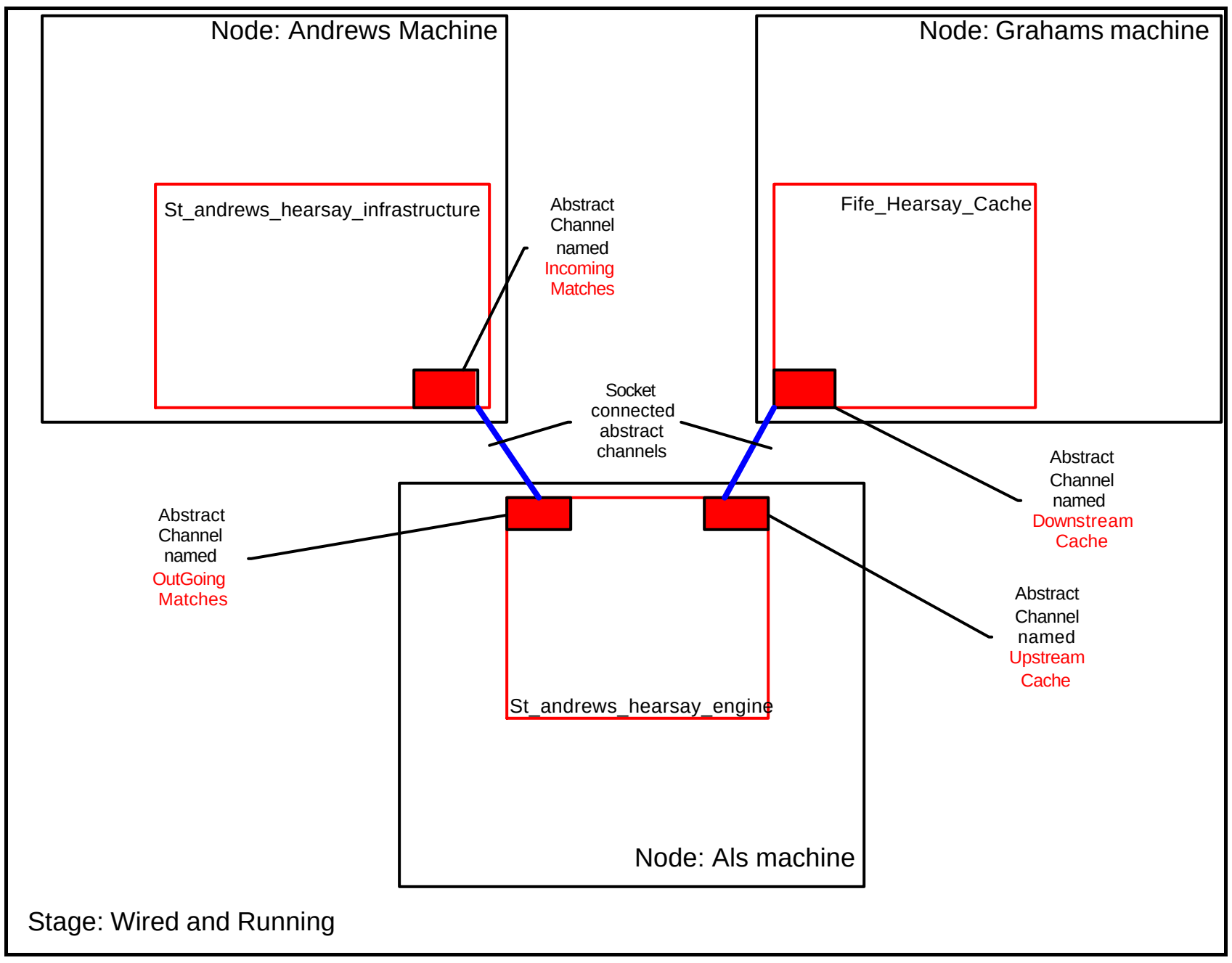

Figure 15: Example of wiring

### 3.4.4 *DETAIL OF WIRING PROCESS*

Wiring takes place once all bundles have been installed and are running. The two nodes which hold the channels to be connected are labeled arbitrarily as the primary and secondary nodes. The primary node is where the wiring process will begin, the other end at which the connection is to be created is known as the secondary node.

One wirer is created and provided with configuration data describing:

1. Locations of both ends of an Abstract Channel to be connected.
2. Service name and provider (binding in PAM) of each running bundle.
3. Name used by executing bundle to reference the channel in both machines (may be different for each machine).
4. Internet Protocol (IP) address of the primary and secondary nodes.

This wirer is sent to the primary node where it communicates with the Connection Manager of the Machine which contains the Abstract Channel to be wired. A socket based protocol is used for the Wirer to communicate with the Connection Manager, the wirer signals that the Connection Manager should create a Server Socket on a free port and when a remote client connects it should bind the connection to a named Abstract Channel.

The wirer now configures another wirer bundle (its 'offspring') which is sent to the secondary node. The purpose of this wirer is to connect the other Abstract Channel on the secondary node to the waiting channel on the primary node. When this wirer arrives at the secondary node, it communicates with the Connection Manager of the machine which requires wiring and instructs it to connect the other Abstract Channel to the Server Socket waiting at the primary node and specified port, the connection is now established.

## 3.5 EVENT MATCHING

Implementation of context-dependent services such as radar, trails and hearsay necessitates complex event processing to detect when user and system contexts match sufficiently to trigger service delivery. This requires a matching engine, a software component that can interpret an input stream of events with respect to a database of matching rules, and take action when matches are detected. The prototype implementations described earlier contain simple specialised matching algorithms. We are also investigating approaches to more general context matching engines.

Events, coming from different systems and sensors form an event cloud, a morass of events which on their own are relatively meaningless but relationships between these events, be they logical, temporal or spatial, may be relatively meaningful. This point forms the basis of a complex event. A complex event can be seen as a virtual event in that it does not actually physically happen within a system but signifies a very real activity based upon the occurrence of other events.

An event pattern language (EPL) describes events, relationships between events and complex events. An event pattern can be seen as a rule base or set of constraints which must be satisfied in order for a complex event to be generated. The power of this language, its flexibility and extensibility are crucial to the success of any complex event processing infrastructure.

A matching engine can be seen as the interpreter of the EPL. This engine takes events as its input and determines matches according to event patterns. This is illustrated in Figure 16.

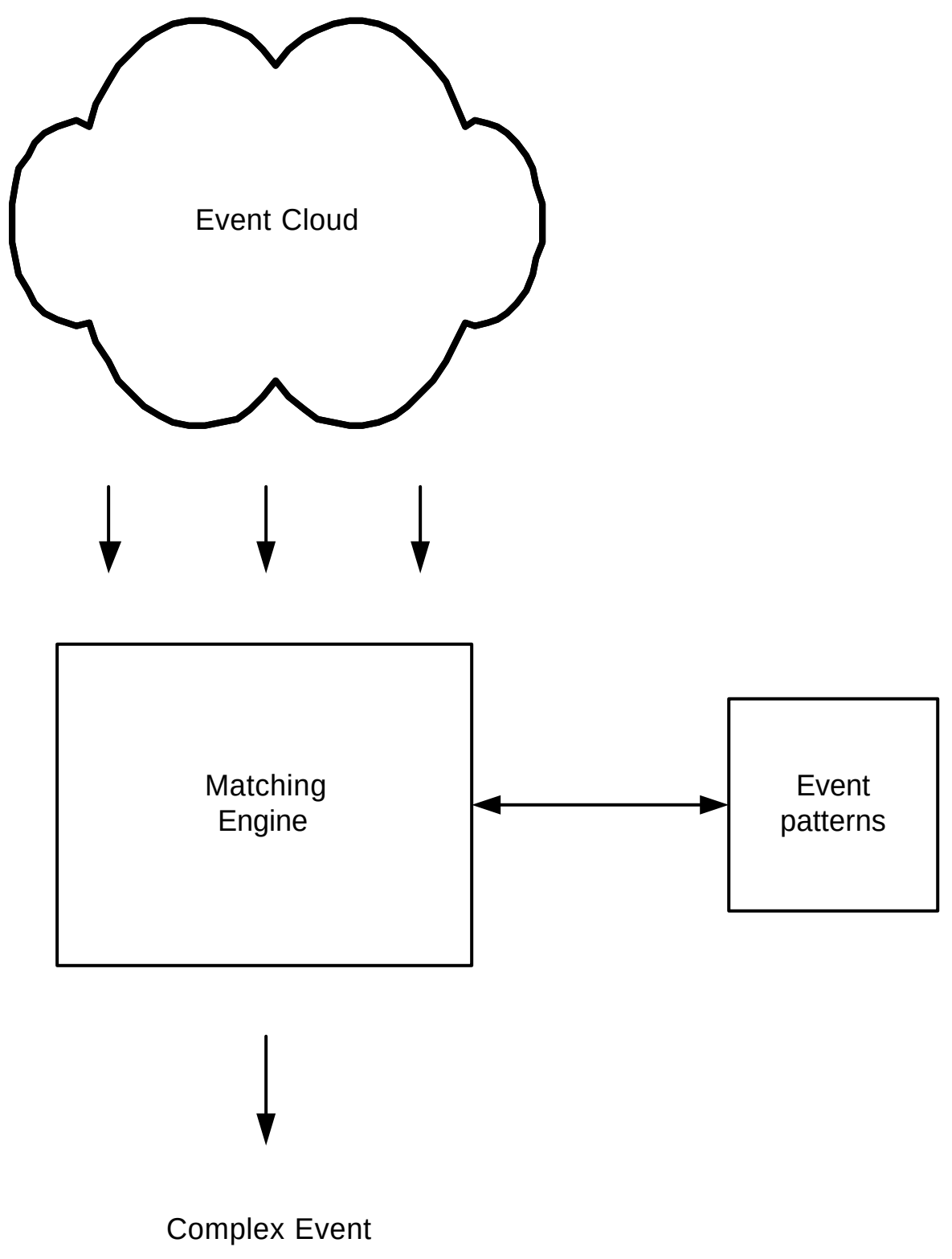

Figure 16: High level design of a Matching Engine

The EPL can be taken further by introducing a means to describe arbitrary computation. This computation, for example, may involve evaluating and updating the state of the matching engine. Statistical computation over sets of events will also be a very useful feature. For example, a complex event may describe the average increase of a particular stock over a given time period.

A number of desirable features for a general matching engine can be identified:

1. An event pattern language (EPL) capable of describing events, relationships between the events and complex events. The EPL must be able to capture logical and temporal relationships as well as causality (the events that cause other events). The EPL should provide such properties as simplicity, expressiveness, rigor and portability.
2. An execution environment within which the EPL is translatable and executable. This environment must be able to take events as input from heterogeneous sources and detect or match events to a given event pattern. The engine must work within the soft real time requirements of the location-aware services it may be applied to.
3. A graphical display allowing administrators to view events coming into the system and the complex events that the engine has generated.
4. Dynamic event pattern evolution, allowing the rules by which matches and complex events are determined to be changed at any time.

We have implemented a prototype hearsay matching service using the Amit matching engine from IBM Research Labs in Haifa. We are currently developing a new matching engine that is tailored to the needs of the Gloss infrastructure.

# 4 PART II – HARDWARE INFRASTRUCTURE (SENSING)

## 4.1 SYSTEM ARCHITECTURE

Figure 17, below, shows a high level view of the Sensor System platform.

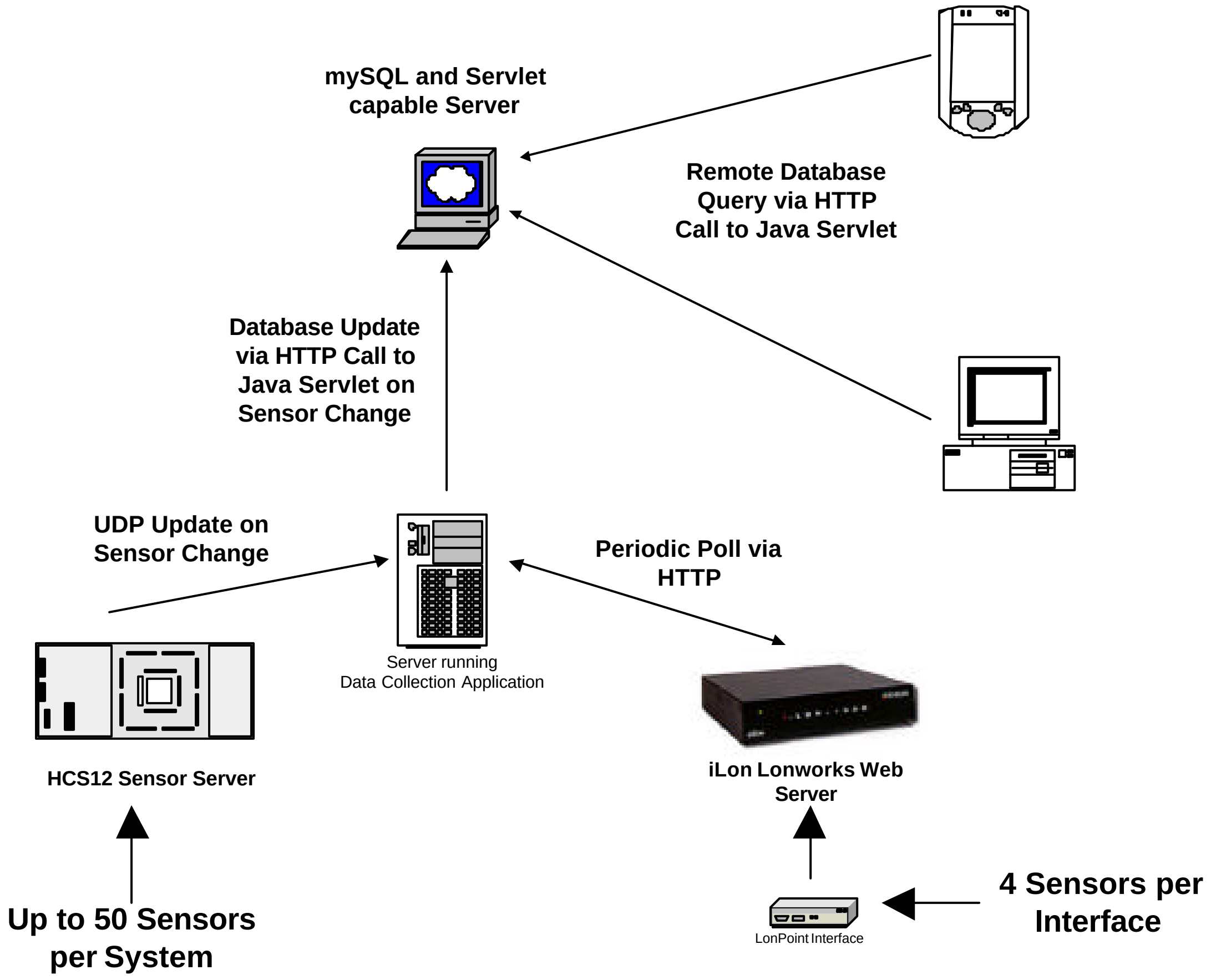

Figure 17: High level view of Sensor System platform

## HCS12 Sensor Server

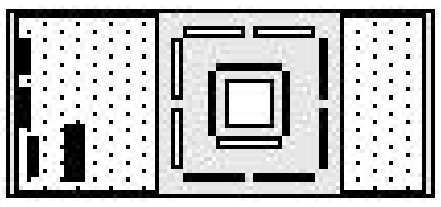

The HCS12 server has been developed as an extension of a project that involved providing network connectivity to the HCS12 general-purpose microcontroller.

Functionality to monitor up to 50 switch-based sensors has been implemented, with sensors being monitored either by HTTP or UDP.

## iLon LonWorks Web Server

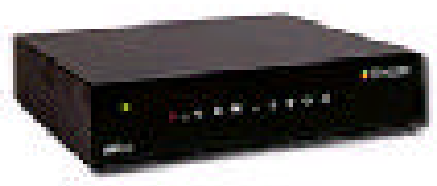

The LonWorks system, developed by Echelon, is designed to provide a peer-to-peer control network for application in building automation and industrial control networks.

The Sensor System only utilises a small part of the facilities provided by LonWorks – the ability to read the values of the sensors connected to the network. This is accomplished by using two types of LonWorks devices, one type to make the current sensor value available (DI-10) and one for providing access to all the readings over a standard data network (iLon 1000 Server). The network data is accessible in the form of Network Variables (NV), which are accessible to any device on the network.

The iLon server also includes an IP interface and on board web server. The web server can be configured to output the state of associated network variables in both HTML and XML formats.

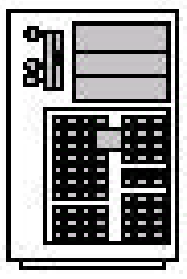

## Data Collection Application

Sensor readings from the iLON and HCS12 servers have to be captured and stored so that the Sensor System can access them at a later date. A persistent data collection application, called DataPull, captures the sensor data from all such devices, formats it, and sends it on to the Servlet to be stored. The DataPull application accesses the iLON and HCS12 servers according to the network protocols they provide. The iLON server is accessed by polling via HTTP, whereas the HCS12 is accessed via UDP and pushes updates to the DataPull application.

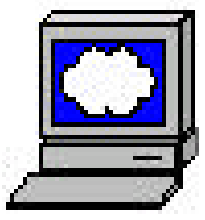

## Servlet and Database Server

The Sensor System Servlet acts as the request broker for the Sensor System platform, as shown in figure 18 below.

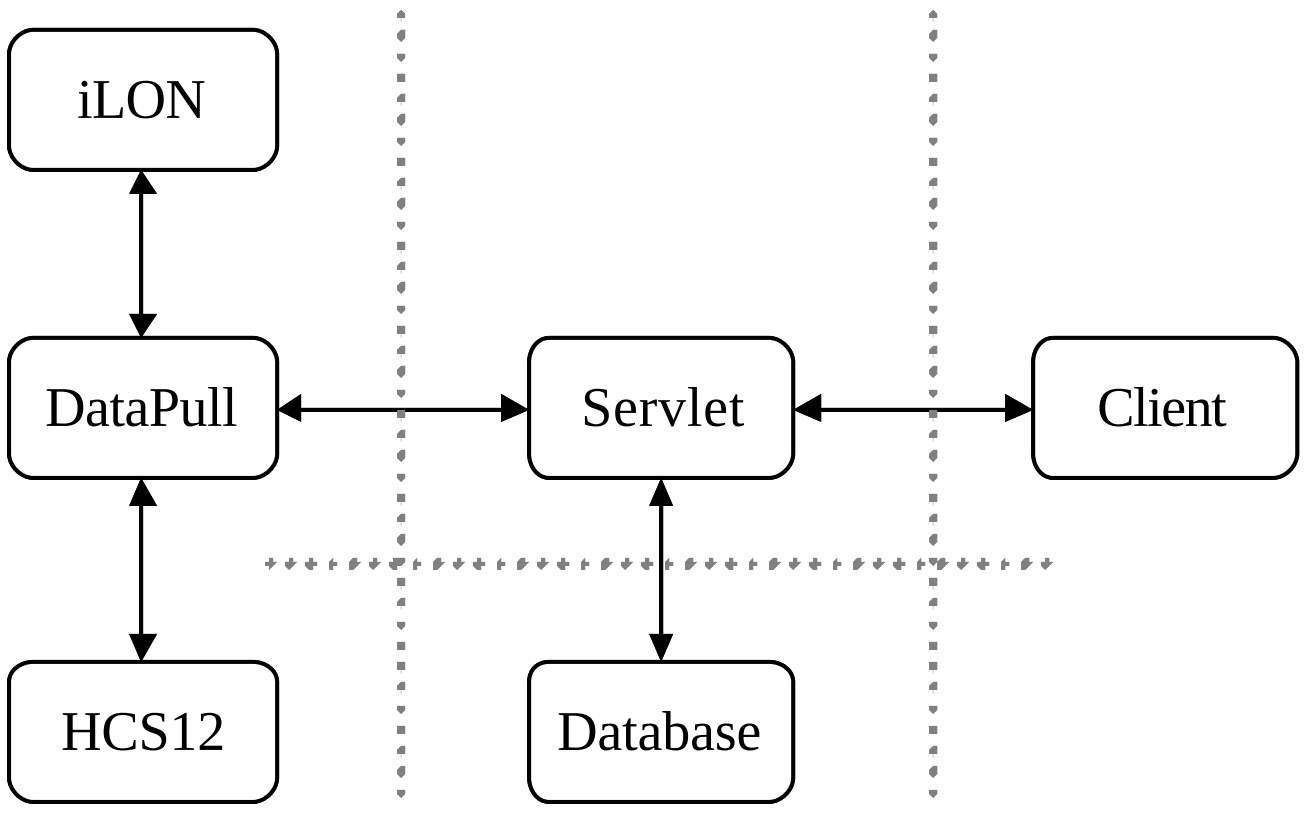

Figure 18: The Servlet is the Object Request Broker (ORB) for the Sensor System platform.

The Sensor System Servlet runs on a Servlet enabled Web server such as Apache Tomcat. The Servlet accepts requests and serves responses via the HTTP protocol. The Servlet also interacts with a remote database, which may or may not reside on the same server, to store sensor data for later retrieval. The use of the Servlet as the request broker for database queries decouples the client-side from the sensor-server-side.

## Clients

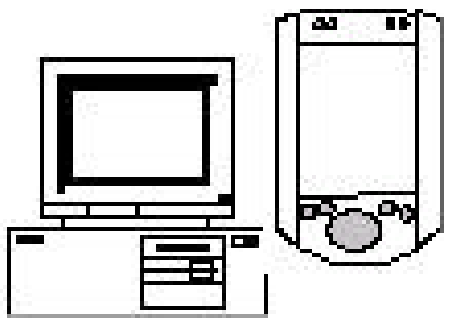

Client applications within the Sensor System platform can access sensor data via the Servlet. The provision of sensor data to clients means that many useful features can be implemented, such as the visualisation of the sensor readings for pattern analysis.

The client application provided as part of the Sensor System platform allows sensor data to be replayed, given a suitable time period, from the sensor readings in the database. It also provides a mode that allows sensor readings to be displayed in real time.

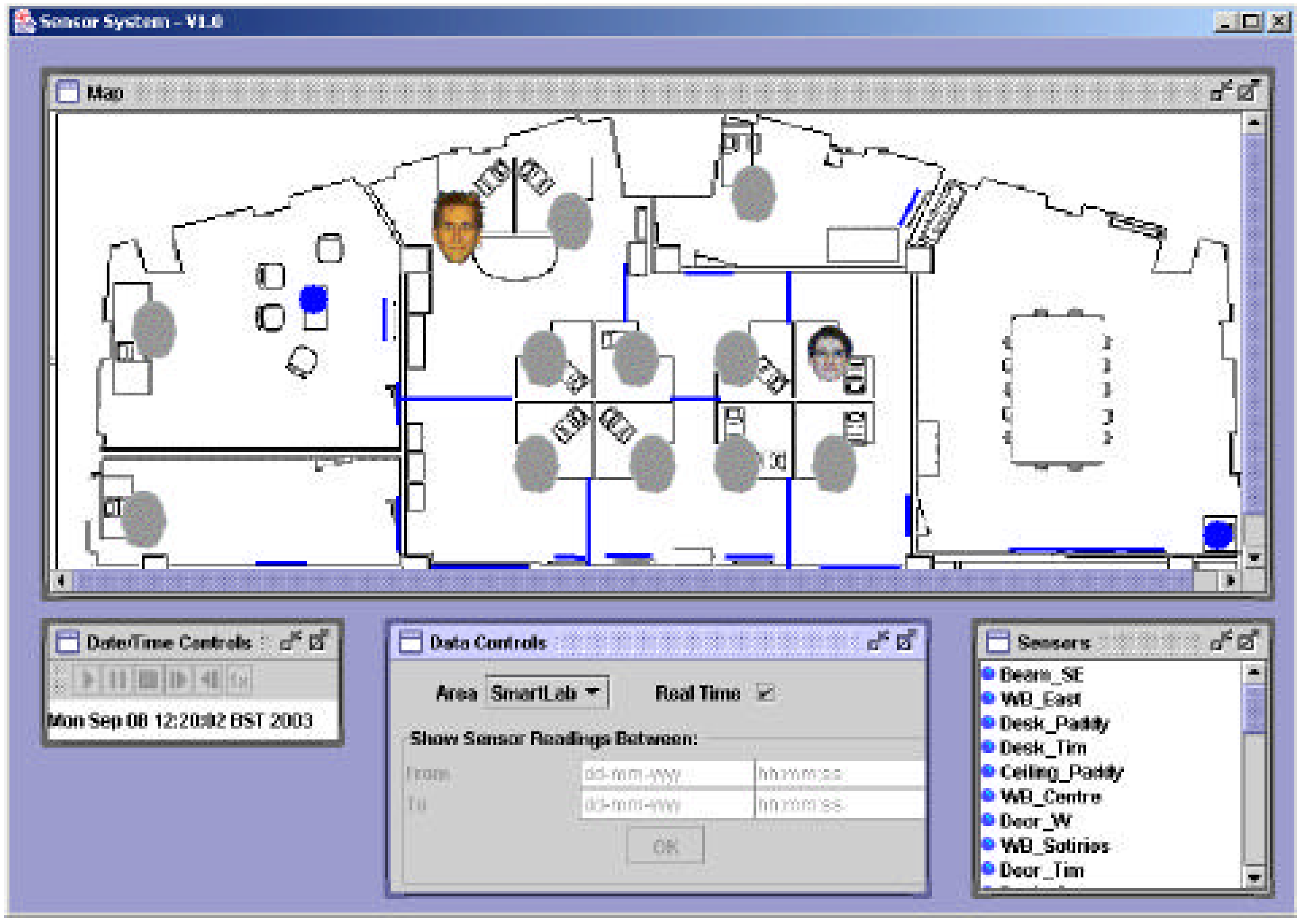

Figure 19: Screenshot of Sensor System Client

# 5 SENSOR HARDWARE CONNECTIONS

This section gives details of the types of sensor hardware used within the Sensor System Platform as well as their physical connection to the sensor servers.

## Sensor Types

A number of different types of sensor have been used, all of which are relatively inexpensive (<£25), anonymous and give an open or closed switch signal.

### PIR (Passive InfraRed)

These detect movement by detecting infrared heat variations in the field of view. The heat emitted from the human body is picked up and electronics inside the sensor look for rapid changes in this reading to differentiate it from false events such as a monitor cooling down after use. The range of the sensors is adjustable, the method varying from brand to brand but is usually between 4 and 12m.

These sensors are utilised throughout the lab for a number of different purposes. Each desk has a PIR underneath to detect the movement of legs that signifies a person sitting at their desk. Every whiteboard has a sensor that detects if someone is writing on the board and the printer has a sensor that detects the presence of someone waiting for a printout.

### Ceiling Mounted Movement Detector

These devices are similar to the PIRs but use microwave radar instead of heat detection. A device of this type is used above the meeting area of Prof. Nixon's office to detect the presence of people sitting in that area.

### Reed Switch

These devices consist of a switch and a magnet. The switch is placed on one section of the object of interest and the magnet on the other. When the switch is in close proximity to the magnet it closes. These switches are commonly used on doors, windows and drawers. In this application they are used to monitor the state (opened/closed) of the six doors in the Smartlab.

### IR Light Beam

The switch on this type of device is controlled by whether or not a reflected light beam is intact. The main unit shines a thin infrared beam towards a reflector, which directs the beam back to the main unit. If there is something blocking the beam then the switch closes.

## Cable Types

Sensors are connected to the sensor servers by different types of cabling. In the case of the Smartlab sensor installation the choice of cable was mainly due to what was available rather than what was most suitable.

Most sensors require 3 wires:

- Power supply (+12V DC). The only exception here is the reed switch sensor, which does not require a power supply as the switch is operated by a magnetic field;
- Ground (GND);
- Input and output signal (for sensor reading). The HCS12 sensors only require one signal wire as a common ground is used within the sensor.

Two types of cable were used to connect the sensors to the servers. These were CAT5 and multi-core alarm cable.

## CAT5 Cable

CAT5 is an Ethernet cable standard defined by the Electronic Industries Association and Telecommunications Industry Association. It contains four pairs of copper wire. CAT5 cable runs are limited to a maximum recommended run rate of 100m (328 feet) for normal 100MB Ethernet connections.

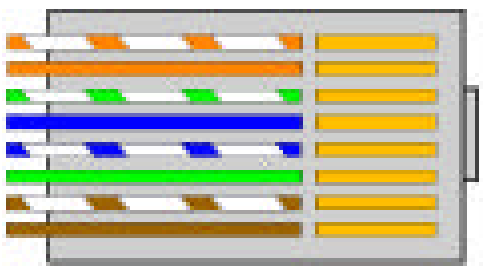

Figure 20: An example of an RJ45 connector showing each of the CAT5 wire pairs

The Sensor System makes use of all of the 8 wires in the CAT5 cable. 6 of the wires are used for sensor signals with the other 2 used to supply power to the sensors. The wiring has been standardised with Orange carrying the +12V DC supply and the Orange/White wire used for Ground. In cases where the sensors have their own power supply all 8 wires are used for sensor signals. Please note that in most cases, RJ45 plugs have not been used, and the Cat5 cable has simply been stripped and wired directly to terminals within sensors and junction boxes.

## Alarm Cable

A mixture of different types of alarm cable was used, again with two cores normally used for power, and remaining cores carrying signal values. Maximum cable run for alarm style cables has not been fully investigated, but runs of around 50m have been used successfully within the department.

# Device connection

## HCS12

All sensors must have a wire returning to the microcontroller, resulting in a star configuration. For ease of wiring, a distributed star configuration can be employed using standard Cat5 cable. The system has been tested successfully over cable runs of up to 30 metres per sensor.

Figure 21, below, shows a central junction box connected to three auxiliary boxes. The configuration below allows up to 18 sensors to be connected to the system and powered from the central point, with each Cat5 cable between the main junction box and auxiliary boxes carrying two power signals and 6 sensor signal wires. If the sensors are sufficiently far away from the controller it may be easier to employ a local power supply enabling 8 sensors to be monitored with a single cat5 cable back to the central point.

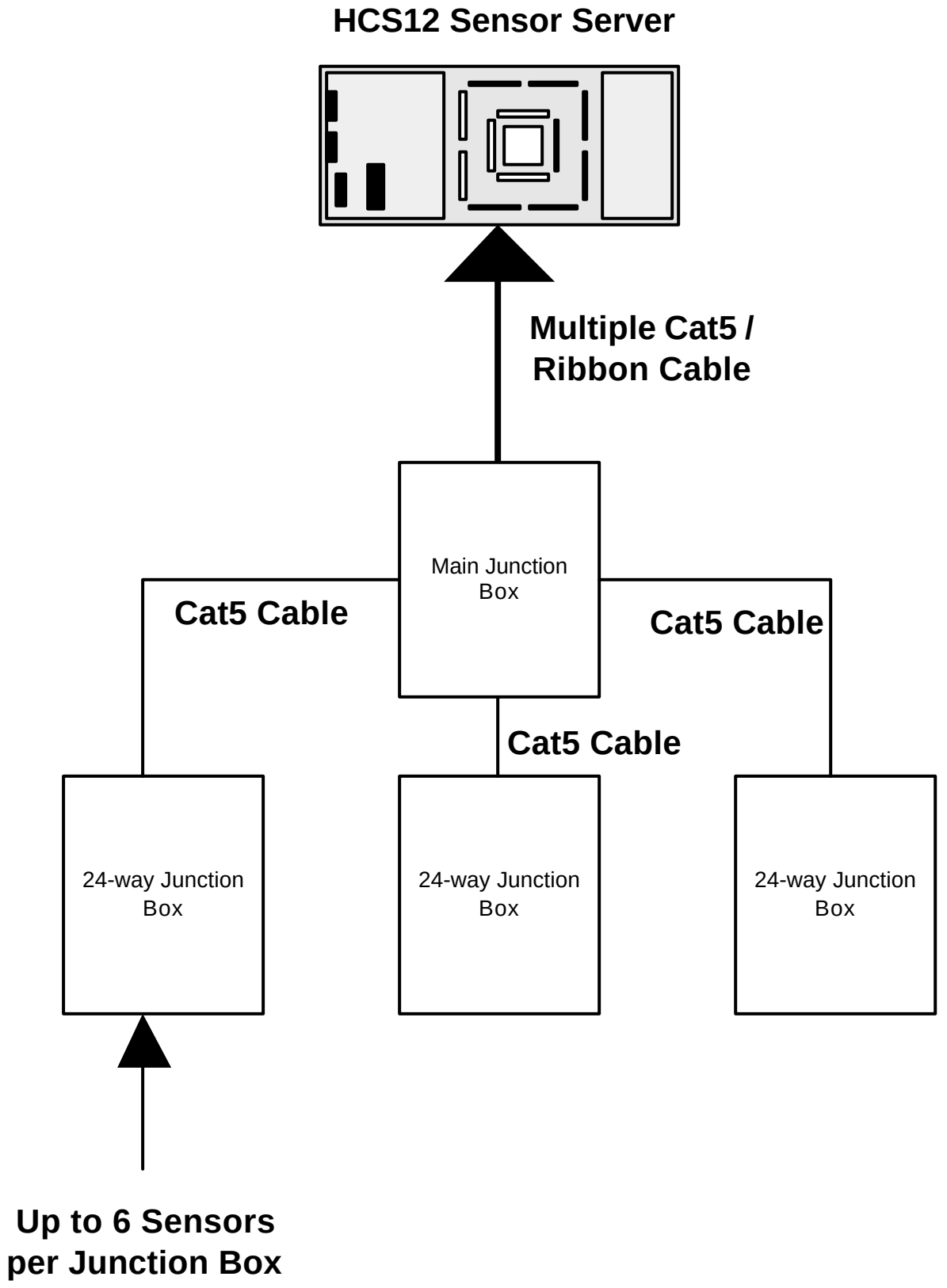

Figure 21: Star configuration of sensor connections to HCS12

If a large number of sensors need to be monitored, a number of HCS12 systems can be deployed, each requiring a separate LAN connection.

## iLon

Sensor connection to the iLon LonWorks based network is slightly different. Each sensor must be connected to a dedicated LonWorks compatible input device by two signal wires. The DI-10 devices used in the department support a maximum of 4 sensor devices each, creating a star configuration at each DI-10 node. Each node must then be connected to both power and a common 2 wire LonWorks channel. If the voltage drop across the channel is small, the whole system can be powered from one supply, with the common bus consisting of two pairs, one for power and one for the LonTalk channel. Again, Cat5 twisted pair can be used as a transmission medium. If the run is too long for one supply, the power pair can be split and a second supply used. Without a dedicated LonWorks bridge, the maximum length of a LonTalk channel is 2700m.

# 6 DATA COLLECTION

This section discusses the format in which data is passed throughout the Sensor System platform so that the sensor readings can be collected and stored in a meaningful way into the database.

## Data Formats

Each part of the Sensor System platform uses a different data format. It was necessary to ensure that the information being passed between parts of the system was minimal to ensure maximum efficiency and robustness of the platform. The Sensor System accumulates a very large volume of data in a short space of time so it was very important that the data interfaces be processed effectively.

### HCS12 to Data Collection Application

The HCS12 sensor server is capable of pushing sensor data to the data collection application when sensor state changes occur. It does this by polling the pins that receive the signals from the sensors. When the server detects a change it creates a UDP datagram packet to send to the data collection application. This datagram packet contains a number of bytes[1], each of which consists of 8 sensor states (1 bit per sensor). The system can tell which sensor is which by determining the bit position being read and then referring to a lookup table to find the ID of the sensor. As the UDP packet overhead is low, each packet contains all sensor states. This builds in some redundancy to cover the unreliable, unacknowledged nature of the UDP protocol.

### iLON to Data Collection Application

The data collection application can retrieve sensor data from the iLon server via HTTP. Unlike the HCS12, which can push sensor data to the data collection application, the iLON server has to be polled for sensor readings. The server maintains an XML document, which holds the sensor names and their associated states. The data collection application regularly polls this XML document and performs a check to determine if any sensor readings have changed since the last time a change was detected. Due to the volume of data traffic being sent across the network and the fact that the data collection application has to perform a comparison every time on all the sensor readings means this method of data retrieval is less effective that the method used by the HCS12.

### Data Collection Application to Servlet

The data collection application sends updates to the sensor readings to the Servlet so that they can be stored into the database. It only sends updates for the sensors that have changed state to avoid wasting network and database resources by sending redundant sensor readings. An XML data format is used between these two applications. The format is simple and includes an entry for each sensor containing the sensor name and new state. The HTTP protocol is used to send the data to the Servlet.

[1] The first byte in the datagram packet is version information.

### Servlet to Clients

Communication between the Servlet and Client applications in the Sensor System platform is via XML request-response messages. The messages sent and received vary according to the context of the query. For example, the request may be for sensor data in real time, or it may be for location-based information about the sensors. Responses from the Servlet may also contain object data representing images for the floor plan of an area containing sensors or for the sensors themselves. Responses from the Servlet to Client applications are sent via HTTP.

More information on the XML data formats used between the Client, Servlet and data collection application can be found in the Servlet External Interface Specification document.

## Data Collection Event Sequencing

This section explains, at a high level, the sequence of events that occur from a sensor being triggered through to the change being logged in the database.

### HCS12 Data Collection

As the system can send UDP packets on sensor change, the DataPull application must listen continuously for these updates. When an update is received, the application checks to see what (if any) sensors have changed from the last update. If changes are detected, these are entered in the database by a HTTP call to the Servlet, which returns an appropriate response.

The sequence diagram, figure 22, below, illustrates the message passing that occurs between the HCS12 system and the database.

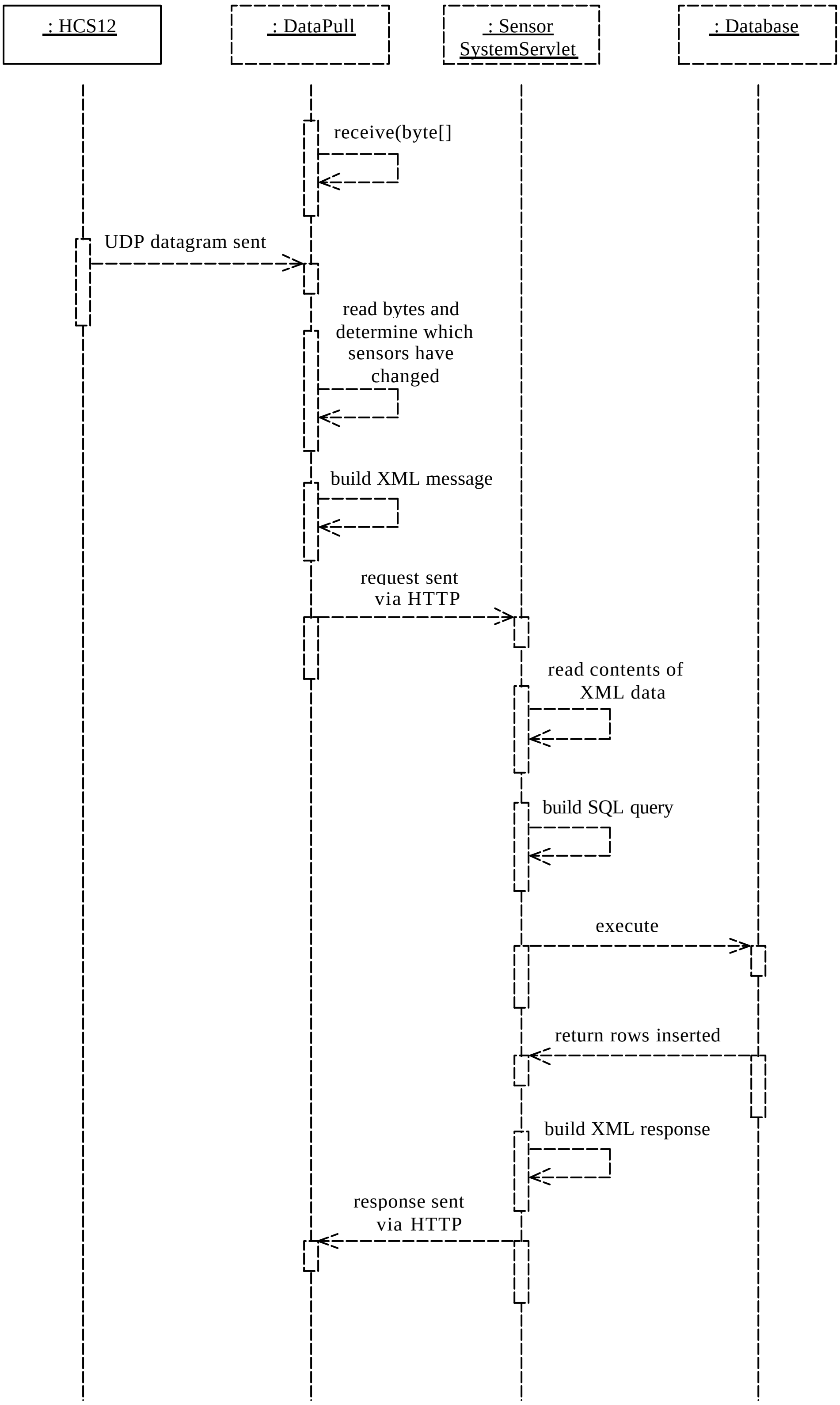

Figure 22: Sequence diagram showing high-level interaction from HCS12 to database

To maintain backwards compatibility, the HCS12 system also contains a web server, and can output sensor states in XML format. This allows the system to use HTTP if, for example, UDP traffic is not carried through a firewall.

# iLON Data Collection

As already discussed, data collection for the iLon system is achieved by a periodic HTTP poll performed by the DataPull application. Once the data has been received and sensor changes extracted, the sequence for updating the database is exactly the same as the case above.

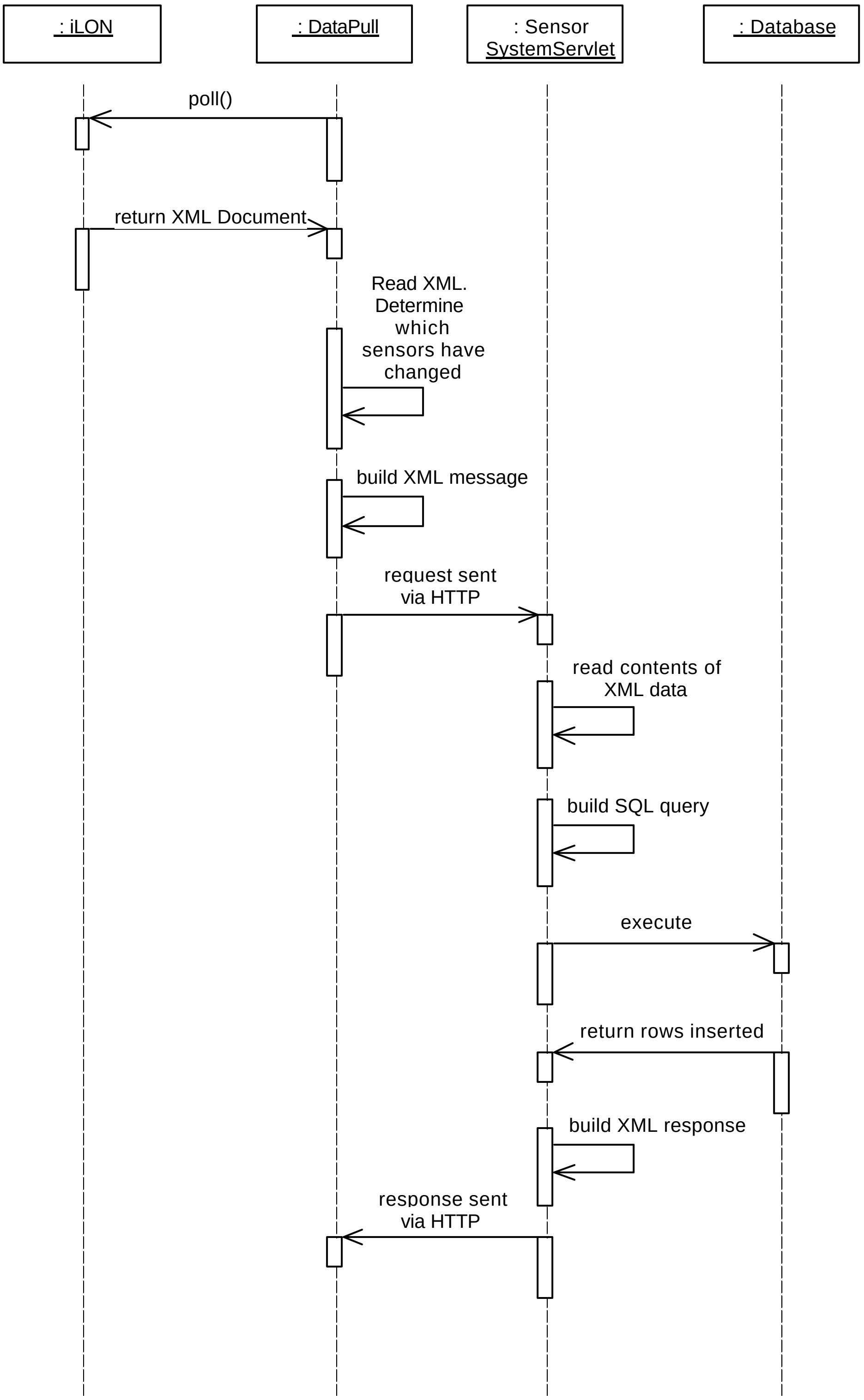

Figure 23: Sequence diagram showing high-level interaction from iLON to database

# 7 DATA PULL APPLICATION

The readings from the multiple servers, whether iLon or HCS12, all have to be centralised and stored in a database, this data collection is performed by the DataPull application. This application listens for UDP packets from any HCS12 devices and polls any iLon devices or HCS12's set up for HTTP.

The application includes a thread which updates the database every second with the latest readings. As the database only has a resolution of a second but some of the sensors can activate and de-activate multiple times within this period, an update will be "activity" if there is any positive (1) readings within the second. The readings are passed to this thread from individual devices through shared hashtables. A database entry is only updated when the reading has changed from the last submission. This results in the database containing the transitions in the state of the sensor rather than a list of every second the sensor was active.

The details of the devices, which are to be polled or listened to by the application, are read from an XML configuration file. Each device is listed as an element with attributes for its configuration and a mapping of sensor names to its inputs. The DTD for the configuration file is given below.

```
<?xml version='1.0' encoding='UTF-8'?>
<!-- DTD for dataPull application configuration file -->
<!ELEMENT SensorConfig (Device+)>
<!ELEMENT Device (Mode)>
<!ATTLIST Device name CDATA #REQUIRED>

<!ELEMENT Mode (Mapping)>
<!ATTLIST Mode protocol (UDP | HTTP) #REQUIRED>
<!ATTLIST Mode address CDATA #IMPLIED>
<!ATTLIST Mode ip CDATA #IMPLIED>
<!ATTLIST Mode port CDATA #IMPLIED>
<!ATTLIST Mode pollsPerSec CDATA #IMPLIED>

<!ELEMENT Mapping (Sensor+)>
<!ELEMENT Sensor EMPTY>
<!ATTLIST Sensor name CDATA #REQUIRED>
<!ATTLIST Sensor inputID ID #REQUIRED>
<!ATTLIST Sensor inverted (true | false) #REQUIRED>
```

For any attributes that are implied, their inclusion is dependant on the protocol being used by the device. The Sensor System configuration tool provides functionality to write this file.

If any HCS12 devices are being used in UDP mode then a listener thread is started to pick up any UDP packets. The listener holds a handler for every device it is listening to and when it receives a packet it directs it to the correct handler according to the IP address. If an IP address is not recognised it is discarded. The device handler extracts the sensor readings from the UDP data and updates the shared hashtable with the results.

For each iLon device or HCS12 in HTTP mode in the system, a thread is started to poll the server and parse the XML to retrieve the sensor readings and add any changes to the shared hashtable. Polling is timed by using a TimerTask, which is run at the required period.

# 8 HIGH LEVEL ARCHITECTURE (CLIENT)

The main functional units that make up the Sensor System Client application are:

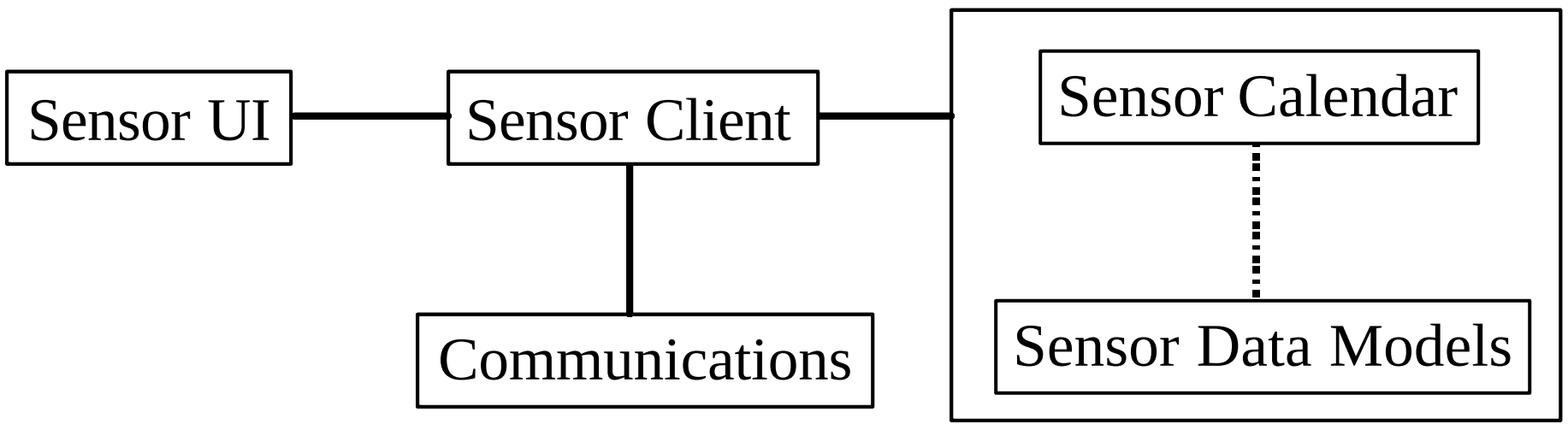

Figure 24: Main functional units of Sensor System Client

The Sensor System Client application was designed such that each functional unit was loosely coupled with the other units. This allows each of the units to be removed and replaced with another one that has similar functionality without having to change the other parts of the application.

## Sensor Client

The Sensor Client is responsible for controlling all of the message passing between the model components, the network and the user interface. Sensor Client maintains the link between all of the components.

## Sensor Calendar

The Sensor System Client's main concern is to show sensor data according to the time at which a sensor state change takes place. Therefore, a calendar component is used to control the flow of data from the underlying data models to any other components that register an interest in the sensor data. This calendar component is navigable, which means that the date and time the calendar is set to can be controlled externally by the user. This allows the data to be displayed in different ways, such as forwards, backwards and at faster playback speeds.

The Sensor Calendar component only plays a role in controlling the data flow for a pre-defined set of data. When the Client requests that real time sensor data is displayed the calendar is no longer navigable because it cannot read data from the future.

## Sensor Data Models

The data models used by the Client are the primary source for obtaining sensor data when it is required to be accessed. Two models exist that reflect the two ways in which the Client can be used to visualise sensor data. These are:

- **Sensor Data Model:** A model that handles historical sensor data and can be used to retrieve the sensor states for a given time period (navigable).

- **Real Time Model:** A model that, once it receives sensor data, will immediately inform any components that have registered an interest in sensor state changes (not navigable).

## Communications

The Sensor System Client obtains all of the sensor information from a remote location accessed via the network. The communications component is responsible for accepting data requests from the Client, sending them to the remote location and then forwarding responses back to the Client.

## Sensor UI

This component represents the graphical front end that will allow the user to interact with the Client, send requests for sensor data and visualise the output from the sensors.

## 8.1 OVERALL CLIENT STRUCTURE

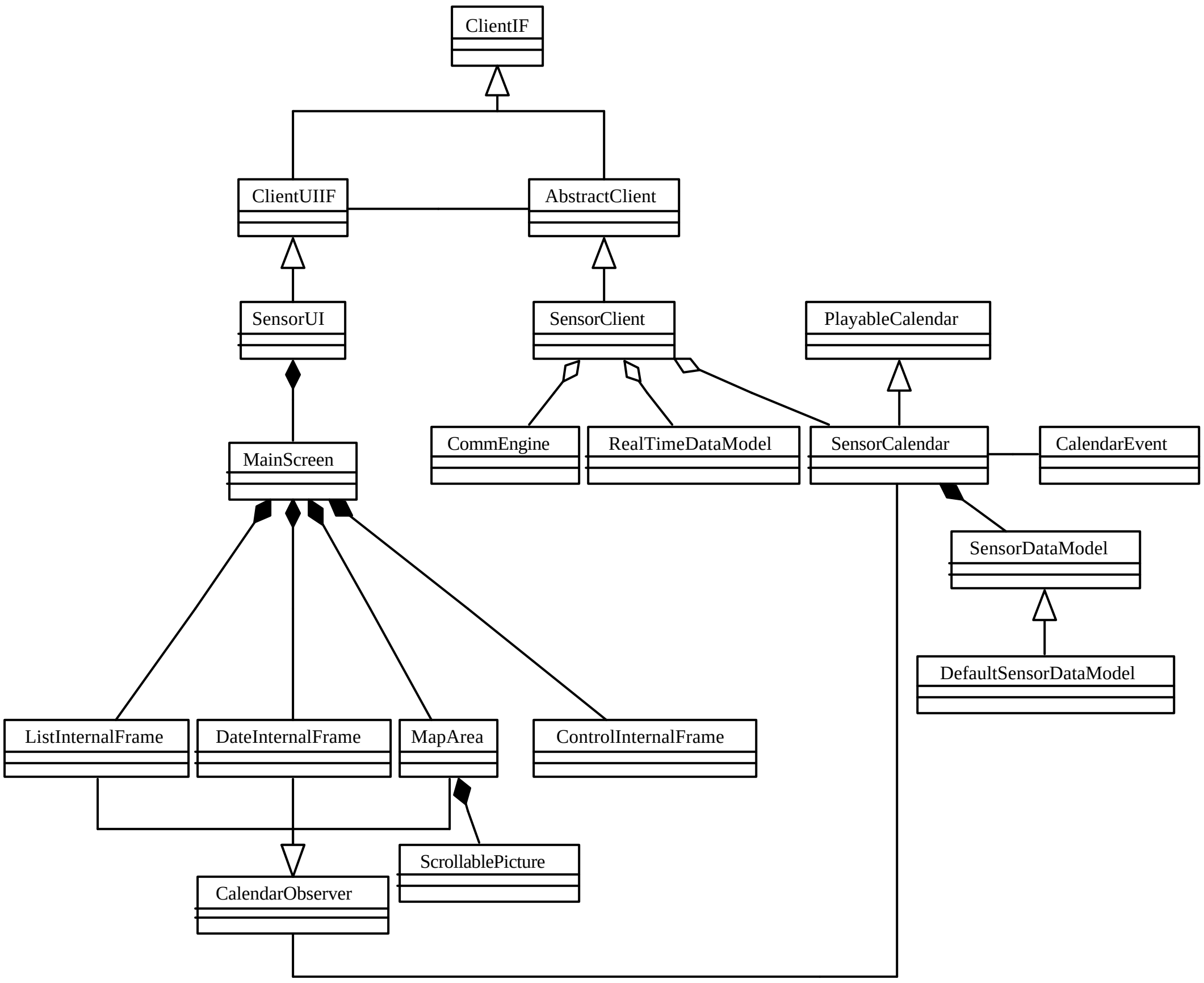

**Figure 25: UML Class Diagram for Sensor System Client**

## 8.2 DETAILED ARCHITECTURE

This section explains some of the design concepts within the Sensor System Client application that require a more detailed explanation.

# Calendar to Data Model Interaction

The sensor data that is made available to the Client application for visualisation purposes comes from a data model. In real time mode the data model is straightforward since it simply notifies the Client of sensor data updates as soon as it receives them. However, when playing back historical sensor data using the calendar and sensor data model the means of obtaining the sensor data is not as straightforward.

## PlayableCalendar and SensorCalendar classes

The main premise of the PlayableCalendar and SensorCalendar classes is that they allow play back, or navigation, of the date and time information held within their internal Calendar objects. This calendar navigation occurs at a pre-defined speed under normal playback conditions, which can be from real time, where the time is updated by 1 second for every 1 second in real time, to a maximum speed where there are many 1 second updates to the time for every 1 second in real time.

Every time the calendar is updated it notifies any observers so that they are aware of the change in time. The calendar will then check for any events that may have occurred at the new time. If an event has occurred then the necessary action is taken and the observers are notified once again but this time they are notified of a calendar event rather than a change in time.

The PlayableCalendar class does not specify any means of checking if an event occurred. The method callTemporalEvent in this class simply updates the calendar time by 1 second and does not perform a check on events. The reason for this is that the PlayableCalendar does not reference any data model from which it may check to determine if any event occurred. It was necessary to create a subclass of PlayableCalendar that referenced a data model. This subclass could then override the callTemporalEvent method to check for data from the data model. The class that does this in the Sensor System is SensorCalendar.

Figure 26, below, shows the operation of the sensor calendar and how it determines if there is new data available for the current calendar time.

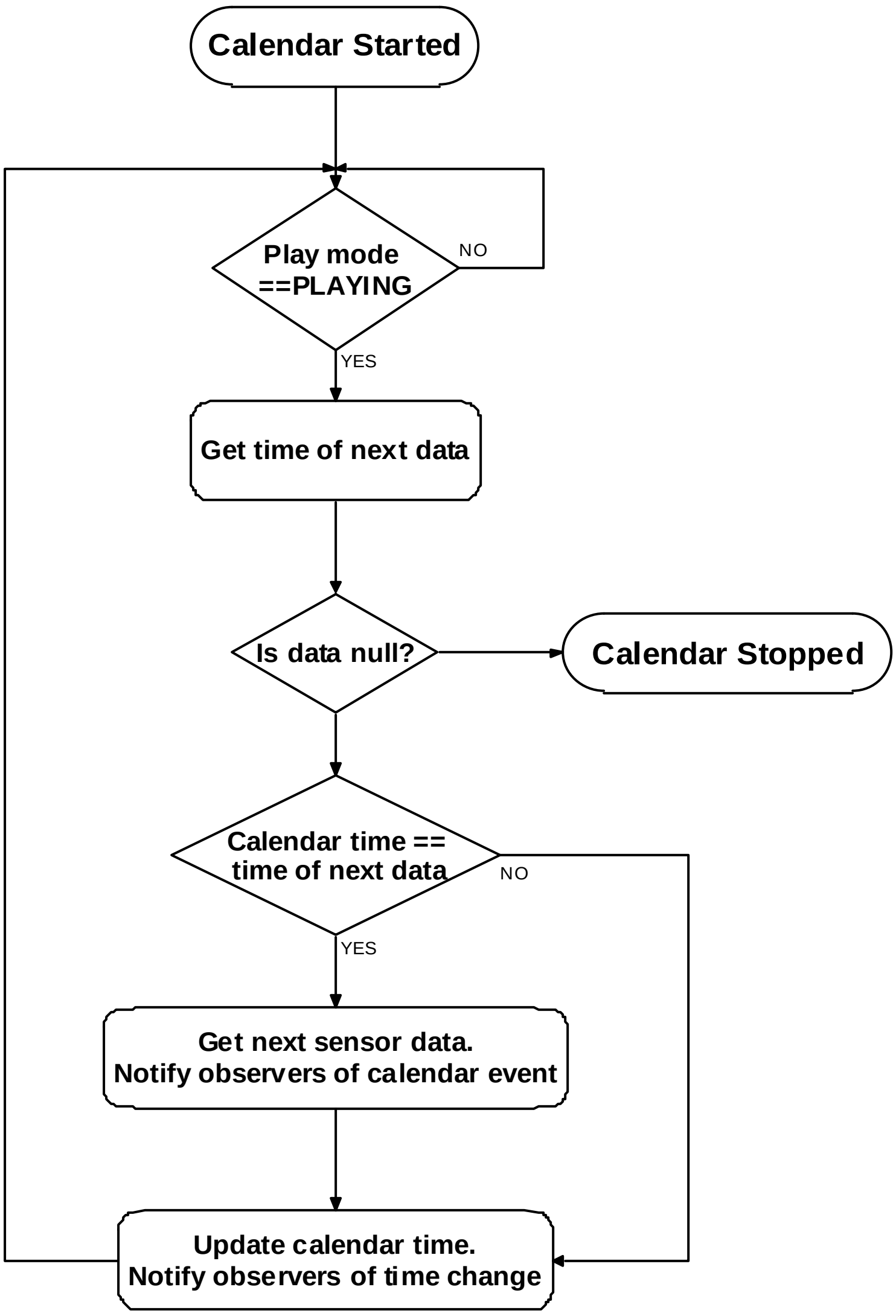

Figure 26: Structure of SensorCalendar main loop

# 9 CONCLUSIONS

The deployment of a suitable software and hardware infrastructure to assess global smart spaces is a development intensive task much more significant than was envisaged in the initial stages. In this document we have shown two aspects of this infrastructure. Firstly, a general software infrastructure for location aware systems that implements the GLOSS architecture detailed in D8. The external location information for this system is supplied by GPS that is translated into a general XML format. The internal location information is provided by a low-cost sensor system developed for the project. We have also described the operational aspects of this system. This system is being packaged for deployment to other partners, but there has been some delay due to limited availability of parts (low power Ethernet cards in particular). The streamed output from the micro web server on the sensor system is tailored to provide location information in the same general XML format. The final stage of work to be undertaken is to deploy, once hardware is available, other instances of the hardware infrastructure at partner sites and undertake a final integration test across multiple geographical locations. This is planned to be between Strathclyde and Trinity given the short timescale.

# 10 APPENDIX 1: WIRING DIAGRAMS

Three areas have been wired using the prototype boards for test purposes. These having been running non-stop now for 2 months. The wiring charts for one of these areas is provided below for reference.

### SMARTLAB WIRING CHARTS

*MAIN JUNCTION BOX*

| Block | Wire | End Sensor | HCS12 Port |
|---|---|---|---|
| 01 | A1 Uplink (G) | Desk_Johnny | S0 (1=movement) |
| 02 | A1 Uplink (G/W) | Desk_Dave | S1 (1=movement) |
| 03 | A1 Uplink (Br) | Desk_Tim | S2 (1=movement) |
| 04 | A1 Uplink (Br/W) | WB_Tim | S3 (1=movement) |
| 05 | A1 Uplink (Bl) | Door_Tim | S4 (1=open) |
| 06 | A1 Uplink (Bl/W) | Beam_NW | S5 (1=clear) |
| | | | |
| 07 | A2 Uplink (G) | Desk_Paddy | S6 (1=movement) |
| 08 | A2 Uplink (G/W) | Ceiling_Paddy | S7 (1=movement) |
| 09 | A2 Uplink (Br) | WB_Paddy | S8 (1=movement) |
| 10 | A2 Uplink (Br/W) | Desk_Sotirios | S9 (1=movement) |
| 11 | A2 Uplink (Bl) | WB_Sotirios | S10 (1=movement) |
| 12 | A2 Uplink (Bl/W) | Door_Paddy | S11 (1=open) |
| | | | |
| 13 | A3 Uplink (G) | Door_E (other door) | S12 (1=open) |
| 14 | A3 Uplink (G/W) | WB_East | S13 (1=open) |
| 15 | A3 Uplink (Br) | Door_Sotirios | S14 (1=movement) |
| 16 | A3 Uplink (Br/W) | Door_W (main door) | S15 (1=open) |
| 17 | A3 Uplink (Bl) | WB_Centre | S16 (1=movement) |
| 18 | A3 Uplink (Bl/W) | WB_West | S17 (1=movement) |
| 19 | A4 Uplink (G) | Beam_SW | S18 (1=clear) |
| 20 | A4 Uplink (G/W) | Desk_Richard | S19 (1=movement) |
| 21 | A4 Uplink (Br) | Desk_Wang | S20 (1=movement) |
| 22 | A4 Uplink (Br/W) | Beam_W | S21 (1=clear) |
| 23 | 0v (Common Ground) | | |
| 24 | +12v dc | | |

*GENERAL NOTES*

+12v dc = **Orange** in Cat5, **Yellow/Green** in 4 core Farnell cable
0v GND = **Orange/White** in Cat5, **Core 1** in 4 core Farnell cable

Sensor Signal **Blue** in Cat5 Cable, use common ground for signal return!

*JUNCTION BOX A1 (NEAR DAVE'S DESK)*

| Block | Wire | End Sensor |
|---|---|---|
| 01 | +12v dc | |
| 02 | +12v dc | |
| 03 | +12v dc | |
| 04 | +12v dc | |
| 05 | +12v dc | |
| 06 | +12v dc | |
| 07 | 0v (Common Ground) | |
| 08 | 0v (Common Ground) | |
| 09 | 0v (Common Ground) | |
| 10 | 0v (Common Ground) | |
| 11 | 0v (Common Ground) | |
| 12 | 0v (Common Ground) | |
| 13 | | |
| 14 | | |
| 15 | | |
| 16 | | |
| 17 | | |
| 18 | | |
| 19 | Uplink (Bl/W) | Beam_NW |
| 20 | Uplink (Bl) | Door_Tim |
| 21 | Uplink (Br/W) | WB_Tim |
| 22 | Uplink (Br) | Desk_Tim |
| 23 | Uplink (G/W) | Desk_Dave |
| 24 | Uplink (G) | Desk_Johnny |

*JUNCTION BOX A2 (IN PADDY'S OFFICE, ABOVE DOOR)*

| Bl ock | Wire | End Sensor |
|---|---|---|
| 01 | Uplink (G) | Desk_Paddy |
| 02 | Uplink (G/W) | Ceiling_Paddy |
| 03 | Uplink (Br) | WB_Paddy |
| 04 | Uplink (Br/W) | Desk_Sotirios |
| 05 | Uplink (Bl) | WB_Sotirios |
| 06 | Uplink (Bl/W) | Door_Paddy |
| 07 | | |
| 08 | | |
| 09 | | |
| 10 | | |
| 11 | | |
| 12 | | |
| 13 | +12v dc | |
| 14 | +12v dc | |
| 15 | +12v dc | |
| 16 | +12v dc | |
| 17 | +12v dc | |
| 18 | +12v dc | |
| 19 | 0v (Common Ground) | |
| 20 | 0v (Common Ground) | |
| 21 | 0v (Common Ground) | |
| 22 | 0v (Common Ground) | |
| 23 | 0v (Common Ground) | |
| 24 | 0v (Common Ground) | |

*JUNCTION BOX A3 (BESIDE MAIN SMARTLAB DOOR)*

| Block | Wire | End Sensor |
|---|---|---|
| 01 | Uplink (G) | Door_W (main door) |
| 02 | Uplink (G/W) | Door_E (other door) |
| 03 | Uplink (Br) | Ceiling_Main |
| 04 | Uplink (Br/W) | Door_Sotirios |
| 05 | Uplink (Bl) | WB_East |
| 06 | Uplink (Bl/W) | WB_West |
| 07 | | |
| 08 | | |
| 09 | | |
| 10 | | |
| 11 | | |
| 12 | | |
| 13 | +12v dc | |
| 14 | +12v dc | |
| 15 | +12v dc | |
| 16 | +12v dc | |
| 17 | +12v dc | |
| 18 | +12v dc | |
| 19 | 0v (Common Ground) | |
| 20 | 0v (Common Ground) | |
| 21 | 0v (Common Ground) | |
| 22 | 0v (Common Ground) | |
| 23 | 0v (Common Ground) | |
| 24 | 0v (Common Ground) | |

*JUNCTION BOX A4 (UNDER FENG WANG DESK)*

| Bl ock | Wire | End Sensor |
|---|---|---|
| 01 | Uplink (G) | Beam_SW |
| 02 | Uplink (G/W) | Desk_Richard |
| 03 | Uplink (Br) | Desk_Wang |
| 04 | Uplink (Br/W) | Beam_W |
| 05 | | |
| 06 | | |
| 07 | | |
| 08 | | |
| 09 | | |
| 10 | | |
| 11 | | |
| 12 | | |
| 13 | +12v dc | |
| 14 | +12v dc | |
| 15 | +12v dc | |
| 16 | +12v dc | |
| 17 | +12v dc | |
| 18 | +12v dc | |
| 19 | 0v (Common Ground) | |
| 20 | 0v (Common Ground) | |
| 21 | 0v (Common Ground) | |
| 22 | 0v (Common Ground) | |
| 23 | 0v (Common Ground) | |
| 24 | 0v (Common Ground) | |